\documentclass[pdflatex,sn-mathphys-ay]{sn-jnl}
\usepackage[latin1]{inputenc}
\usepackage[T1]{fontenc}
\usepackage{amsmath,bm}
\usepackage{amsfonts}
\usepackage{amssymb}
\usepackage{graphicx}
\usepackage{caption}
\usepackage{mathtools}
\usepackage{multirow}%
\usepackage{amsthm}%
\usepackage{mathrsfs}%
\usepackage{textcomp}%
\usepackage{booktabs}%
\usepackage{listings}%
\usepackage{enumitem}%

\begin{document}

\title{Pair Space Analysis of Some Four-Body Central Configurations}

\author*[1]{\fnm{Alon} \sur{Drory}}\email{alond@afeka.ac.il}

\affil*[1]{\orgname{Afeka College of Engineering}, \city{Tel-Aviv}, \country{Israel}}

\maketitle

\begin{abstract}
	
I study central configurations in the four-body problem, i.e., configurations in which the forces on all the bodies point to a fixed, single point in space. The newly formulated pair-space formalism yields a set of vectorial equations that fully characterize such configurations. I investigate a sub-class of solutions in which at least two pairs of inter-body distances are equal. The only such non-collinear configurations are the tetrahedron (the unique non-planar configuration), kites and the isosceles trapezium. The specific shapes (internal angles) are determined by the ratio of the masses of the bodies. Mathematical expressions are given for all these relations.
\end{abstract}


\section{Introduction}
\label{sec:intro}

Assume a system containing $N$ particles with masses $\{m_i\}$ and vector positions $\{ \bm{r}_i\}$ taken in geometrical space $\mathbb{R}^3$. In many cases, from planetary systems to gas molecules, the particles are interacting via pairwise potentials $v_{ij}( \bm{q}_{ij} )$, where $\bm{q}_{ij} = \bm{r}_i - \bm{r}_j$. The difficulty in such systems lies in the coupling between the equations introduced by the relative positions $\left\{\bm{q}_{ij}\right\}$. In previous works \citep{Drory a, Drory b}, I suggested representing such classical mechanical systems in a new space, henceforth called pair-space, which is spanned by the system's center of mass $\bm{R}$ and the pair positions $\{\bm{q}_{1 2},..., \bm{q}_{(N-1) N}\}$. 

Effectively, pair space considers pairs of particles as the fundamental objects rather than the particles themselves. Therefore, the system's potential energy is decoupled, being a sum of independent terms
\begin{equation}
	V = \sum\limits_{[ij]} v_{ij}(\bm{q}_{ij})   ,
\end{equation}
where $\sum_{[i,j]}$ means a sum over all pairs of distinct indices with $i < j$.

In pair space, the system's kinetic energy is 
\begin{equation}
	\label{kinetic}
	T = \frac{1}{2}M\bm{\dot{R}}^2 + \sum_{[i,j]}\frac{1}{2}\mu_{ij}\bm{\dot{q}_{ij}}^2 - \sum_{[i,j,k]}\frac{1}{2}\mu_{ijk}\left(\bm{\dot{q}_{ij}} + \bm{\dot{q}_{jk}} + \bm{\dot{q}_{ki}}\right)^2   ,
\end{equation}
where $M=\sum_{i=1}^N m_i$ is the system's total mass and $\sum_{[i,j,k]}$ is a sum over all triplets of distinct indices such that $i < j < k$. $\mu_{ij}$ and $\mu_{ijk}$ are the pair and triplet reduced masses, respectively, defined as
\begin{subequations}\label{reduced:main}
	\begin{align}
		\mu_{ij} &= \dfrac{m_i m_j}{M}   ,\label{reduced:a}\\  
		\mu_{ijk} &=\dfrac{m_i m_j m_k}{M^2} .\label{reduced:b}
	\end{align}
\end{subequations}
 
Hamilton's principle does not apply to the Lagrangian $L = T-V$ because the pair positions are not all independent, as they verify the so-called ``triangle conditions''
\begin{equation}
	\label{triangle}
	\bm{q}_{ij} + \bm{q}_{j k} + \bm{q}_{k i} = 0 .
\end{equation}
A configuration $\{\bm{q}_{i j} \}$ that verifies these conditions will be termed ``\textbf{realizable}''. 

Rather than use them to reduce the number of variables, we treat Eqs.(\ref{triangle}) as \textit{dynamical} constraints. To this end, we introduce vector Lagrange multipliers, $\bm{\phi}_{ijk}$, defined for every triplet of monotonically ordered distinct indices. For convenience, we also define formal symbols where the ordering is different by 
\begin{equation}
	\label{sgn}
	\bm{\phi}_{\sigma(i) \sigma(j) \sigma(k)} = sgn(\sigma)\bm{\phi}_{i j k}   ,
\end{equation}
where $\sigma$ is a permutation of the indices $(i j k)$.

Hamilton's principle and the Euler-Lagrange equations hold for the pair Lagrangian, $L_{\pi}$, defined as
 \begin{equation}
 	\label{lagrangianp}
 	L_{\pi} = T  - V + \sum_{[i,j,k]}\bm{\phi}_{ijk} \left( \bm{q}_{ij} + \bm{q}_{j k} + \bm{q}_{k i}\right) .
 \end{equation}
 
After some manipulation of the Euler-Lagrange equations, we can write the equations of motion in the following form \citep{Drory a}:

\begin{equation}
	\label{eqmotionq}
	\mu_{ij} \bm{\ddot{q}}_{ij}  +\frac{\partial v_{ij}(\bm{q}_{ij})}{\partial \bm{q}_{ij}} - \bm{J}_{ij} = 0  ,
\end{equation}
for any pair of indices $i < j$, where we defined
\begin{eqnarray}
	\label{jij}
	\bm{J}_{ij} = \sum_{\substack{n=1 \\n\neq i,j}}^{N}  \bm{\phi}_{ijn}   .
\end{eqnarray}
Note that $\bm{J}_{ji}=- \bm{J}_{ij}$. 

The explicit formula for these terms is \citep[see][]{Drory a},
\begin{subequations} \label{eqmotionJ}
	\begin{align}
		&\frac{1}{\mu_{ij}}\bm{J}_{ij} = \sum_{\substack{k = 1 \\ k \neq i, j}}^{N} \dfrac{m_k}{M} \bm{F}_{ijk} , \label{eqmotionJ:a}\\
		\text{where}  \nonumber \\
		&\bm{F}_{ijk} = 
		\frac{1}{\mu_{ij}}\frac{\partial v_{ij}(\bm{q}_{ij})}{\partial \bm{q}_{ij}} + \frac{1}{\mu_{jk}}\frac{\partial v_{jk}(\bm{q}_{jk})}{\partial \bm{q}_{jk}} + \frac{1}{\mu_{ki}}\frac{\partial v_{ki}(\bm{q}_{ki})}{\partial \bm{q}_{ki}} .  \label{eqmotionJ:b} 
	\end{align}
\end{subequations}

In particular, for the Newtonian potential, $v_{ij}(q_{ij}) = - \dfrac{G M \mu_{ij}}{q_{ij}}$, we have that 
\begin{equation}
	\bm{F}_{ijk} = G M \left( \frac{\bm{q}_{ij}}{q_{ij}^3} + \frac{\bm{q}_{jk}}{q_{jk}^3} + \frac{\bm{q}_{ki}}{q_{ki}^3}  \right)  .
\end{equation}
This is the basis for the famous gravitational N-body problem, with which the present work is dealing.

This paper continues the exploration of the pair space formalism begun in \citet{Drory a} and \citet{Drory b}. Because this is a new formalism, at this stage we concentrate on showing how it functions as a unifying and simplifying framework. Consequently, many results presented here are known already, but they were derived separately, with ad hoc assumptions or methods, whereas our aim here is to show how the pair space representation offers a methodical approach, from which previously disparate results follow as mere particular cases, encompassed in a consistent and general approach. 

In previous papers we presented a new characterization of central configurations, which are a particular but important subset of N-body solutions. We continue to discuss these configurations here, in the context of the four-body problem, which is the simplest N-body problem for which not all these configurations are known, although many specific central configurations have been investigated before \citep{Albouy et. al., Alvarez-Ramirez and Llibre, Corbera et. al. b, Corbera et. al., Deng and Zhang, Dziobek, Erdi and Czirjak, Kovari and Erdi, Macmillan and Bartky, Xie}. 

The problem of enumerating all possible four-body central configurations is too complex to be solved completely here. In accordance with the aims of the present study, we concentrate on showing how the pair-representation offers a compact, fully-vectorial, systematic and methodical approach to the investigation of these configurations, rather than attempt exhaustiveness. Consequently, we concentrate on the simplest solutions. Specifically, we shall assume that some pairs of relative distances are equal in the configuration. As we shall see, this implies that the configurations possess some kind of symmetry. We do not assume anything about the masses, however, in contrast to most previous studies. The conditions on the masses will arise naturally out of the equations themselves. 

The main results of this study are as follows. I stress once again that most of these results are known, but derived independently and by different approaches and methods. The aim of the present work is to show how pair-space offers a systematic method for investigating all these cases within a unified framework.

1. The only spatial (non-planar) configuration is the regular tetrahedron, for which all the pair-distances are equal. There is no restriction on the masses, so that any body may have any arbitrary mass.

2. A planar configuration can have at most three equal pair-distances. In this case, one body is equidistant from all the others. This body must then lie at the centre of an equilateral triangle formed by the other three bodies. These latter bodies must all have identical masses. The mass of the body at the centre of the triangle is arbitrary.

3. If there is a body that is equidistant from only two other bodies, rather than three, the configuration is a kite. Two bodies lie at the ends of the symmetry axis, and two others are placed in mirror positions at equal distances from the axis. These last two masses must be equal. The ratio of the masses at the ends of the symmetry axis, on the other hand, is determined by the internal angles that fix the shape of the kite. These angles fall within restricted ranges of values, which depend on whether the kite is concave or convex.

4. Configurations where no body is equidistant from at least two other bodies may still possess and axis of symmetry. The central configuration is then an isosceles trapezium. The bodies that lie along each of the bases must have equal masses. The ratio of the mass values of bodies that belong to different bases is a function of the base angle of the trapezium, i.e., it is determined by the configuration's shape.

This paper is therefore structured as follows: section \ref{sec:central} presents the pair space representation of central configurations, then in section \ref{sec:4bodcent}, we focus on the central configurations of the four body problem. The form of the equations will then imply why the simplest solutions are those where we assume that some pair distances are equal. These will be the focus of the remainder of the paper. Section \ref{sec:spatial} shows why most central configurations are planar by presenting a simple proof that there is only one possible non-planar four-body central configuration. Among the planar cases, an extreme case is configurations that are not only planar but collinear. These occupy section \ref{sec:collinearity}. For planar but non-collinear configurations, section \ref{sec:massindep} presents some purely geometrical constraints, which are mass-independent, and generalize the well-known Dziobek relations \citep{Dziobek}. The remaining sections then derive specific shapes. Sections \ref{sec:equilateral} and \ref{sec:isosceles} analyze the kite configurations, while section \ref{sec:parallel} discusses the parallelogram and isosceles trapezium cases. Further details on the aim and structure of the present analysis can be found at the end of section \ref{sec:4bodcent}.

\section{Central configurations}
\label{sec:central}

Since pairs of particles are the fundamental entities in pair space, we can define angular momenta for each one. The pair angular momentum of $[i,j]$ is defined as
\begin{equation}
	\bm{L}_{ij}= \bm{q}_{ij} \times \mu_{ij}\bm{\dot{q}}_{ij}  .
\end{equation}
The total pair angular-momentum, $\sum_{[i,j]}\bm{L}_{ij}$ is always conserved (this is equivalent to the conservation of the system's usual total angular momentum) \citep{Drory a, Drory b}. Usually, however, individual pair angular momenta will not be conserved. Solutions that do conserve all individual pair momenta exist, and have important properties, because such solutions are central configurations \citep{Drory b}. 

A central configuration is defined as one in which the acceleration of each particle point towards the system's center of mass, so that 
\begin{equation*}
	\ddot{\bm{r}}_i = - \lambda \left( \bm{r}_i - \bm{R} \right) ,
\end{equation*}
where $\lambda$ is identical for all indices $i$. In pair space, this is equivalent to 
\begin{equation}
	\label{defcc}
	\bm{\ddot{q}}_{i j} = - \lambda \bm{q}_{i j}   .
\end{equation}

Central configurations hold special importance in the study of the $N$-body problem \citep{Moeckel, Saari b, Saari a, Hampton and Moeckel}. The Lagrange and Euler solutions for three bodies, for example, are central configurations. More specifically, they are homographic, meaning that they are configurations that remain self-similar at all times. Pizzeti proved that homographic solutions are central at all times \citep{Pizzeti}. 

Another subtype of central configurations are those that display homothetic motion. This means that the bodies converge along fixed straight lines until they collide. Homothetic solutions are also homographic, and thus form central configurations at all times. Not all $N$-body collisions arise from homothetic motions, however. Configurations that lead to $N$-body collisions non-homothetically are not central; nevertheless, as the bodies approach each other, their orbits will asymptotically approach a central configuration \citep{Siegel and Moser}. Conversely, when a system expands and its bodies separate from each other, it also tends towards a central configuration \citep{Saari a}. Other properties of central configurations and the fundamental role they seem to play in the solution of the $N$-body problem can be found in \citet{Moeckel, Saari b, Saari a}.

It follows from Eq.(\ref{defcc}) that in a central configuration, the associated pair-momentum is conserved for every pair, since:
\begin{equation}
	\dfrac{d \bm{L}_{i j}}{dt} = \bm{q}_{ij} \times \mu_{ij}\bm{\ddot{q}}_{ij} = 0  .
\end{equation}
Hence, if a solution is a central configuration at all times, all its pair angular momenta are conserved individually. In \citet{Drory b}, I have shown that the reverse also holds. Thus central configurations can be characterized as precisely those that conserve all individual pair angular momenta. This yields an algebraic characterization of these configurations.

As shown in \citet{Drory b}, a \textit{realizable} configuration that does not remain on a single fixed line at all times is central if and only if, for every pair $(i,j)$:
\begin{equation}
	\label{cceq}
	\sum_{\substack{k = 1 \\ k \neq i, j}}^{N}  m_k \left(  \bm{q}_{ij} \times \bm{q}_{jk} \right)  \left(\frac{1}{q_{ik}^3} - \frac{1}{q_{jk}^3} \right) = 0
\end{equation}

The condition ``realizable'' is necessary since it may be that some algebraic solutions of these equations have no geometrical meaning. For example, it is clear that if all the mutual distances $q_{ij}$ are equal to each other, Eqs.(\ref{cceq}) hold trivially. But geometrically, such a configuration represents a regular simplex with $N$ vertices, a structure that cannot exist in a space of dimension less than $N-1$. In our three-dimensional space, therefore, such a solution can exist only for $N \leq 4$, even though Eqs.(\ref{cceq}) are solved by any regular simplex configuration, independently of the value of $N$. Thus, the solutions of Eqs.(\ref{cceq}) must be checked to make sure that they are actually realizable solutions of the equations of motion.

If the configuration is planar, all the vector products in Eq.(\ref{cceq}) are perpendicular to the plane of the motion. They can then be replaced by the oriented area of the parallelogram built on the pair-position vectors. This scalar version of Eq.(\ref{cceq}) was first derived by Ernesto Laura in the specific context of the motion of N vortices \citep{Laura}, and independently for the general N-body problem, by Henri Andoyer \citep{Andoyer}. The Laura-Andoyer equations have been popularized by Hagihara in \citet{Hagihara} an have been used extensively since \citep{Moeckel, Albouy et. al., pinalong, pina, Fernandes}. Meyer seems to have written these equations in higher dimensions in \citet{meyer}, chapter I, but still referring to planar configurations. A generalization of his work can be found for example in \citet{hampton}.

It is worth noting, however, that Eqs.(\ref{cceq}), apart from being a vectorial generalization of the Laura-Andoyer equations, are derived from a conservation principle, namely the conservation of all pair angular-momenta, which illuminates their origin from an interesting physical point of view.

\section{Four-body central configurations}
\label{sec:4bodcent}

Let us now consider the central configuration equations in the case of four bodies. In Eqs.(\ref{cceq}), we can manipulate the vector products so as to have only four different expressions. Arbitrarily, I have chosen to use $\bm{q}_{21}  \times \bm{q}_{31}$, $\bm{q}_{21}\times \bm{q}_{41}$, $\bm{q} _{41}  \times \bm{q}_{31}$, $\bm{q}_{42}  \times \bm{q}_{32}$. Other products can be reduced to these by using the triangle condition, e.g., $\bm{q} _{13}  \times \bm{q}_{23} = -\bm{q}_{31}  \times \left( -\bm{q}_{31} - \bm{q}_{12}\right) = \bm{q}_{21}  \times \bm{q}_{31}$ and so on. Eqs.(\ref{cceq}) become a set of six equations (one for each pair of bodies):

\begin{subequations} \label{4cc}
	\begin{align}
m_3 \left[ \dfrac{1}{q_{31}^3} - \dfrac{1}{q_{32}^3}\right]\left( \bm{q} _{21}  \times \bm{q}_{31}\right) &= m_4 \left[ \dfrac{1}{q_{42}^3} - \dfrac{1}{q_{41}^3}\right]\left( \bm{q}_{21}  \times \bm{q}_{41}\right)          
\label{4cc:a}\\
m_2 \left[ \dfrac{1}{q_{21}^3} - \dfrac{1}{q_{32}^3}\right]\left( \bm{q}_{21}  \times \bm{q}_{31}\right) &= m_4 \left[ \dfrac{1}{q_{43}^3} - \dfrac{1}{q_{41}^3}\right]\left( \bm{q}_{41}  \times \bm{q}_{31}\right)           
\label{4cc:b}\\
m_1 \left[ \dfrac{1}{q_{21}^3} - \dfrac{1}{q_{31}^3}\right]\left( \bm{q} _{21}  \times \bm{q}_{31}\right) &= m_4 \left[ \dfrac{1}{q_{42}^3} - \dfrac{1}{q_{43}^3}\right]\left( \bm{q}_{42}  \times \bm{q}_{32}\right)            
\label{4cc:c}\\
m_2 \left[ \dfrac{1}{q_{42}^3} - \dfrac{1}{q_{21}^3}\right]\left( \bm{q} _{21}  \times \bm{q}_{41}\right) &= m_3 \left[ \dfrac{1}{q_{43}^3} - \dfrac{1}{q_{31}^3}\right]\left( \bm{q} _{41}  \times \bm{q}_{31}\right) 
\label{4cc:d}\\
m_1 \left[ \dfrac{1}{q_{41}^3} - \dfrac{1}{q_{21}^3}\right]\left( \bm{q}_{21}  \times \bm{q}_{41}\right) &= m_3 \left[ \dfrac{1}{q_{23}^3} - \dfrac{1}{q_{43}^3}\right]\left( \bm{q}_{42}  \times \bm{q}_{32}\right) 
\label{4cc:e}\\
m_1 \left[ \dfrac{1}{q_{41}^3} - \dfrac{1}{q_{31}^3}\right]\left( \bm{q}_{41}  \times \bm{q}_{31}\right) &= m_2 \left[ \dfrac{1}{q_{32}^3} - \dfrac{1}{q_{42}^3}\right]\left( \bm{q}_{42}  \times \bm{q}_{32}\right)
\label{4cc:f}
	\end{align}
\end{subequations}

There are many solutions to these equations and trying to find them all would be far beyond the scope of this paper. Instead, as mentioned in section \ref{sec:intro}, we shall concentrate on the simplest solutions. A look at the structure of the equations shows that they greatly simplify when some relative distances are equal. In such a case, terms of the form $\left[ \dfrac{1}{q_{ij}^3} - \dfrac{1}{q_{jk}^3}\right]$ vanish, which in turn implies the vanishing of the other side of the equation in which these terms appear. Configurations of this type turn out to have high symmetry and as such have been investigated before \citep{Macmillan and Bartky, Albouy et. al., Alvarez-Ramirez and Llibre,  Corbera et. al. b, Deng and Zhang, Kovari and Erdi, Xie}. 

Most investigations assume some combination of a particular geometry, conditions on the masses and a specific coordinate system tailored to the specific configuration investigated (e.g., \citet{Kovari and Erdi, Shoaib}). Admittedly, the questions they investigate are difficult and thus some restrictions on the type of configuration considered is usually required. Thus, for example, \citet{Alvarez-Ramirez and Llibre} assume both a certain symmetry and conditions on the masses, with the aim of counting exactly the number of central configurations of a given type. \citet{Xie} assumes both a particular symmetry and mass equalities to tackle the genuinely difficult problem of uniqueness of trapezoidal central configurations. On the other hand, \citet{Albouy et. al.} prove that the configuration's symmetry implies some relations between the masses. Similar theorems have been proved by \citet{LongSun, DengLi, PerezSanto, FernandesLlibre}.

It is worth nothing, however that these investigations typically assume coordinates chosen specifically for the symmetry assumed or proved. Furthermore, because each geometry is treated as a separate case, the mathematical representation of the equations, i.e., the coordinates used, are chosen ad-hoc for every type of configuration. By contrast, Eqs.(\ref{4cc}) are fully vectorial, and do not rely on a specific choice of coordinates. In fact, at no point are we introducing specific coordinates into the equations. The results are purely geometric, relying on distances and angles, obtained directly from the vectorial form of the equations. As we shall see, this allows for the derivation of many properties from vector operations and identities, thereby unifying the various cases and simplifying their treatment. We also do not assume anything about the masses beforehand. The equations themselves generate conditions on the masses once geometrical assumptions are made. 

 \section{Spatial configuration}
\label{sec:spatial}

First, we examine which configurations are non-planar (spatial). 

\textbf{Proposition 1:} The only four-body non-planar (spatial) configuration is the regular tetrahedron.

\textbf{Proof:} 

Let us take the scalar product of Eq.(\ref{4cc:a}) with $\bm{q}_{31}$. The left-hand side vanishes and we obtain the identity
\begin{equation}
\left[ \dfrac{1}{q_{41}^3} - \dfrac{1}{q_{42}^3}\right]\left( \bm{q} _{41}  \times \bm{q}_{21}\right) \cdot \bm{q}_{31} = 0   .
\end{equation}
Therefore, either $q_{41} = q_{42}$, or $\bm{q}_{41} , \bm{q}_{31} , \bm{q}_{21} $ are all coplanar. If the latter, then all four bodies are coplanar. This means that any non-planar solution must have $q_{41} = q_{42}$. By taking scalar products of the other equations with the appropriate vectors, we can show that any non-planar solution must have that $q_{21} = q_{31} = q_{41} = q_{32} = q_{42} = q_{43}$. This means the configuration is a regular simplex with four vertices, i.e., a tetrahedron. \textbf{QED}

As mentioned above, this is the highest $N$ allowable for this type of configuration, since regular simplices with more vertices can only exist in spaces of higher dimension.

The tetrahedral solution is the four-body analog of Lagrange's three-body solution, in the following sense: as in the Lagrange solution, every pair of bodies moves in a two-body Keplerian orbit. This follows from the fact that for any triplet of distinct indices $(i , j , k)$, we have from Eq.(\ref{eqmotionJ:b}) that if $q_{ij} = q_{k l}$ for every pair of indices, then
\begin{equation}
	\bm{F}_{ijk} = GM\left[ \dfrac{\bm{q}_{i j}}{q_{i j}^3} + \dfrac{\bm{q}_{j k}}{q_{j k}^3} + \dfrac{\bm{q}_{k i}}{q_{k i}^3}\right] = \dfrac{G M}{q_{i j}^3}\left[\bm{q}_{i j} + \bm{q}_{j k} +\bm{q}_{k i}\right] = 0  .
\end{equation}
Eq.(\ref{eqmotionJ:a}) shows that $J_{i j} = 0$ for every pair of indices, and thus, from Eq.(\ref{eqmotionq})
\begin{equation}
	\label{4bodq}
\mu_{ij} \bm{\ddot{q}}_{ij}  +\frac{G M \mu_{ij}}{q_{ij}^3} \bm{q}_{ij} = 0   ,
\end{equation}
which is the Keplerian equation of motion.

The existence of the tetrahedral solution has been known for a long time \citep{Lehmann}.  The types of possible dynamics are much more limited here than in the three-body case, however. In the Lagrange solution, each pair of bodies may perform any two-body Keplerian motion, whereas here, it is known \citep{Wintner} that a spatial central configuration with $N > 3$ can only perform homothetic motions, i.e., all bodies collapse onto each other or they all escape from each other.

\section{Possible Collinearity}
\label{sec:collinearity}

Any configuration that is not tetrahedral must be planar. An extreme case is when some masses are collinear. It turns out, however, that we cannot have a configuration that is only partially collinear.

\textbf{Proposition 2:} If any three masses are collinear, they all are.

\textbf{Proof:}

Assume that three bodies are collinear. Without loss of generality, let us label them $m_1, m_2, m_3$ . The collinearity implies that the vector product $\bm{q}_{21} \times \bm{q}_{3 1} = 0$. From Eq.(\ref{4cc:a}), either $\bm{q}_{4 1} \times \bm{q}_{2 1} = 0$ or $q_{4 1}  = q_{4 2}$. If the former, then all bodies are collinear and the proof is concluded. Let us therefore assume that $q_{4 1}  = q_{4 2}$. 

In Eq.(\ref{4cc:b}), the vector product in the left-hand-side vanishes, hence 
\begin{equation*}
	\left[ \dfrac{1}{q_{43}^3} - \dfrac{1}{q_{41}^3}\right]\left( \bm{q} _{41}  \times \bm{q}_{31}\right) = 0   .
\end{equation*}
Again, if the cross product were to vanish, all the bodies would be collinear. If that is not the case, we must have that $q_{4 3} = q_{4 1}$.

Thus, either all four masses are collinear, or else $q_{4 1} = q_{4 2} = q_{4 3}$. The latter possibility makes $m_1, m_2, m_3$ equidistant from $m_4$, which implies that they must lie on the circumference of a circle centered on $m_4$ . But then they cannot be collinear, in contradiction to the assumption. Hence, if one vector product vanishes (and three bodies are collinear), they must all be collinear. \textbf{QED}

The properties of the collinear solution of $N$ bodies have been analyzed in \citet{Drory b} and I shall not repeat them here. For the remainder of this paper, we therefore assume that our configurations are planar and non-collinear.

\section{Mass-independent relations}
\label{sec:massindep}

As shown in \citet{Drory b}, Eqs.(\ref{4cc}) imply a set of geometric constraints that are independent of the masses of the bodies. Any configuration whose shape violates these relation cannot be a central configuration; hence, they are useful to potentially eliminate possible configuration shapes from consideration as central configurations. Since the general treatment given in \citet{Drory b} is somewhat unwieldy, I give here an alternative derivation, more direct and streamlined. 

To simplify slightly the notation, let us introduce the symbols
\begin{equation}
	p_{i j} = p_{j i} = \dfrac{1}{q^3_{i j}}  .
\end{equation}
Every triplet of masses generates a relation through the equations in which only these masses appear. For example, only the masses $m_1, m_3 , m_4$, appear in Eqs.(\ref{4cc:a}), (\ref{4cc:c}) and (\ref{4cc:e}). To eliminate them, form the following scalar products:
\begin{eqnarray*}
	\text{I} = (p_{24}-p_{34})(p_{23}-p_{34})\cdot \text{Eq.(\ref{4cc:a})}\cdotp(\textbf{q}_{42} \times \textbf{q}_{32}) \\
	\text{II} = (p_{23}-p_{34})(p_{24}-p_{14})\cdot \text{Eq.(\ref{4cc:c})}\cdotp(\textbf{q}_{21} \times \textbf{q}_{41}) \\
	\text{III} = (p_{13}-p_{23})(p_{24}-p_{34})\cdot \text{Eq.(\ref{4cc:e})}\cdotp(\textbf{q}_{21} \times \textbf{q}_{31})
\end{eqnarray*}
Now subtract $\text{II}$ from the sum of  $\text{I}$ and $\text{III}$ (i.e., $\text{I} + \text{III} - \text{II}$). This yields
\begin{equation*}
	m_1(\textbf{q}_{21} \times \textbf{q}_{31})\cdotp(\textbf{q}_{21} \times \textbf{q}_{41}) \cdot \mathfrak{B} = 0    ,
\end{equation*}
where
\begin{equation*}
	\mathfrak{B} = (p_{13}-p_{23})(p_{24}-p_{34})(p_{14}-p_{12}) -(p_{24}-p_{14})(p_{12}-p_{13})(p_{23}-p_{34})    .
\end{equation*}
Since the configuration is non-collinear, none of the vector products vanishes, and since it is planar, these vector products are all normal to the configuration's plane, hence none of the scalar products vanishes. Thus we must have $\mathfrak{B} = 0$. Repeating this procedure for the other three triplets of masses, we obtain four mass-independent relations that every planar, non-collinear central configuration must verify:
\begin{subequations} \label{dziobek}
	\begin{align}
(p_{13}-p_{23})(p_{24}-p_{34})(p_{14}-p_{12}) &=(p_{24}-p_{14})(p_{12}-p_{13})(p_{23}-p_{34}) \label{dziobek:a} \\
(p_{12}-p_{23})(p_{13}-p_{34})(p_{14}-p_{24}) &=(p_{12}-p_{24})(p_{13}-p_{23})(p_{14}-p_{34}) \label{dziobek:b} \\
(p_{13}-p_{14})(p_{23}-p_{12})(p_{34}-p_{24}) &=(p_{13}-p_{12})(p_{23}-p_{24})(p_{34}-p_{14}) \label{dziobek:c} \\
(p_{14}-p_{13})(p_{24}-p_{12})(p_{34}-p_{23}) &=(p_{14}-p_{12})(p_{24}-p_{23})(p_{34}-p_{13}) \label{dziobek:d} 
	\end{align}
\end{subequations}
	
Eq.(\ref{dziobek:a}) is equivalent to the well known Dziobek relation \citep{Dziobek, Corbera et. al.}, derived already in 1900, but here we have obtained three more. These relations are not always all independent. For example, if all the mutual distances are different, so that no side of the equations vanishes, multiplying Eq.(\ref{dziobek:a}) by (\ref{dziobek:b}) and dividing the result by Eq.(\ref{dziobek:c}) will yield Eq.(\ref{dziobek:d}). 

On the other hand, there is more information here than just the Dziobeck relation, Eq.(\ref{dziobek:a}). Consider for example the possibility that $p_{12} = p_{13} = p_{23}$. Eq.(\ref{dziobek:a}) is verified automatically, along with Eqs.(\ref{dziobek:b}) and (\ref{dziobek:c}), but Eq.(\ref{dziobek:d}) is not, hence it contains additional restrictions in this case. 

Note that these relations are necessary conditions for a configuration to be central, but not sufficient, as they do not restrict the values of the masses. Eqs.(\ref{4cc}), on the other hand, are sufficient, and thus one should rely primarily on them. However, the mass-independent relations suggest, once more, that the equality of mutual distances is an important criterion. We shall therefore consider the implications of such equalities now.

\section{Body-centered equilateral triangle}
\label{sec:equilateral}

We start with the assumption that there are at least three equal distances between bodies so that they form an equilateral triangle. We shall prove the following 

\textbf{Proposition 3:} If three bodies in a central configuration form an equilateral triangle, then

(a) The masses of the three bodies must all be equal.

(b) The fourth body must lie at the center of the triangle and its mass is arbitrary.

\textbf{Proof:}

\begin{figure}[h!]
	\includegraphics[width=0.7\linewidth]{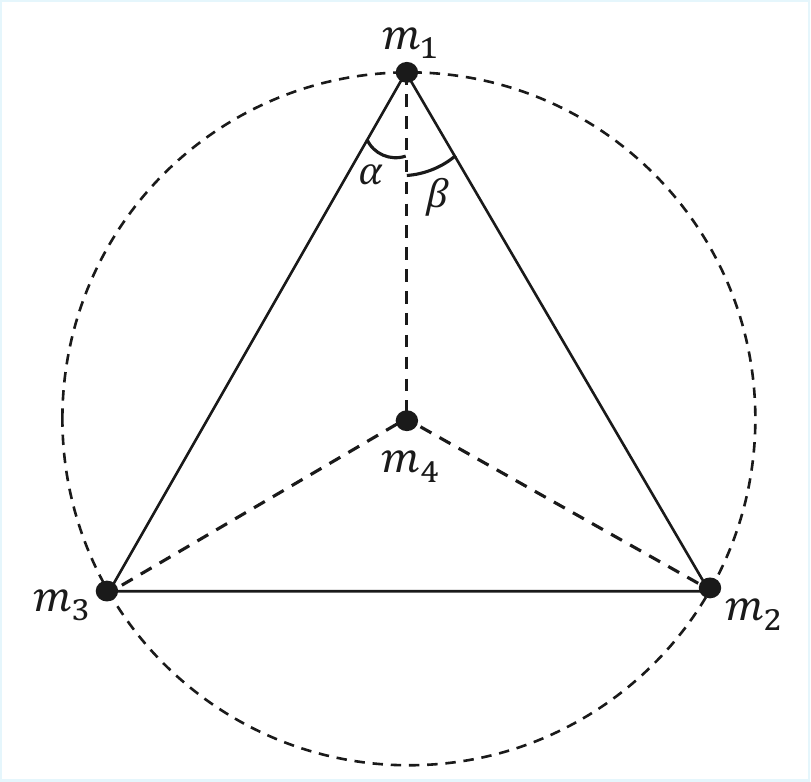}
	\caption{The bodies $m_1 , m_2 , m_3$ are the vertices of an equilateral triangle inscribed in a circle centered on $m_4$}
	\label{Fig:4body_equilateral}
\end{figure}

Without loss of generality, label the three masses that form the equilateral triangle as $m_1 , m_2 , m_3$. Since $\triangle 123$ is equilateral, we have that $q_{21} = q_{23} = q_{31}$. From Eq.(\ref{4cc:a}), $\left[ \dfrac{1}{q_{41}^3} - \dfrac{1}{q_{42}^3}\right]\left( \bm{q} _{41}  \times \bm{q}_{21}\right) = 0$. Since the cross product cannot vanish, we must have that $q_{41}= q_{42}$. By the same argument, Eq.(\ref{4cc:b}) implies that $q_{41} = q_{43}$. Hence, $m_4$ is equidistant from all the other masses. If we draw a circle centered on $m_4$, with radius $q_{41}$, the other three masses lie on the circle. The equilateral triangle $\triangle 1 2 3$ is inscribed in this circle, and therefore $m_4$ must be inside the triangle, see Fig.\ref{Fig:4body_equilateral}.

Next, we use Eqs.(\ref{4cc}) to prove that $m_1 = m_2 = m_3$. In Eq.(\ref{4cc:d}), the factors $\left[ \dfrac{1}{q_{21}^3} - \dfrac{1}{q_{42}^3}\right]$ and $\left[ \dfrac{1}{q_{43}^3} - \dfrac{1}{q_{13}^3}\right]$ cancel out, and similarly in Eqs.(\ref{4cc:e}) and (\ref{4cc:f}). We obtain the three relations (only two of which are independent):
\begin{subequations}
	\label{m2m3}
		\begin{align}
	m_2 \left( \bm{q}_{21}  \times \bm{q}_{41}\right) &= m_3 \left( \bm{q}_{41}  \times \bm{q}_{31}\right) \label{m2m3:a} \\
	m_1 \left( \bm{q}_{21}  \times \bm{q}_{41}\right) &= m_3 \left( \bm{q}_{42}  \times \bm{q}_{32}\right) \label{m2m3:b} \\
	m_1 \left( \bm{q}_{41}  \times \bm{q}_{31}\right) &= m_2 \left( \bm{q}_{42}  \times \bm{q}_{32}\right) \label{m2m3:c} 
		\end{align}
\end{subequations}
Note that the order of the factors in the vector products has been altered to simplify the signs. Taking the norm of Eq.(\ref{m2m3:a}) yields
\begin{equation}
	m_2 q_{41} q_{21} \sin \beta = m_3 q_{41}q_{31}\sin \alpha 
\end{equation}
where $\alpha = \measuredangle 314$ and $\beta = \measuredangle 412$. From Fig.1, $\alpha = \beta = \pi /6$, thus $m_2 = m_3$. Similarly from Eq.(\ref{m2m3:b}), we have that $m_1 = m_2 = m_3$. The central mass $m_4$ remains arbitrary. This completes the proof of the proposition.

That this configuration is realizable, i.e., that there exists a solution of the equations of motion that retains this shape, up to rotations and rescaling, is a classic result which can be found in, e.g., \citet{Hagihara} or \citet{Wintner}. Nevertheless, for completeness sake, I have included a proof of this that uses the pair-space formalism, in the Appendix. The aim was to show how pair-space methods are applicable to proofs of this type. The proofs of realizability of other cases discussed in the present work follow similar lines. Since these results are also well known, I do not include separate proofs for every type of configuration analyzed here.

\section{Kites}
\label{sec:isosceles}

We now assume a weaker symmetry, namely, that among the bodies, there are three that form an isosceles triangle that is not equilateral. We show that

\textbf{Proposition 4:} 

(a) Such central configurations must have the shape of a kite.

(b)  The masses of the two bodies that lie on either side of the symmetry axis are equal.

\textbf{Proof:}

Without loss of generality, we label the mass at the apex as $m_1$, and the other two as $m_2$ and $m_3$. Thus, $q_{21} = q_{31}$. From Eq.(\ref{4cc:c}),we have again that $q_{43}= q_{42}$. Hence, the triangle $\triangle 243$ is also isosceles and the bodies form a kite (either concave or convex). These conditions suffice to verify all the mass-independent relations, Eqs.(\ref{dziobek}), which thus hold no additional information.

There are three cases, depending on the position of $m_4$. One case yields a convex kite (see Fig.\ref{Fig:convex_kite}) and two cases yield concave kites (see Figs.\ref{Fig:concave1} and \ref{Fig:concave2}). In all cases, we denote by $\alpha$ the angle between $\bm{q}_{12}$ and $\bm{q}_{23}$, i.e., the base angle. Similarly, $\beta$ is always the base angle of $\triangle 243$, defined between $\bm{q}_{24}$ and $\bm{q}_{23}$. Once we know whether the kite is convex or concave these two angles define its shape up to scaling. Note that $\alpha, \beta < \pi / 2$, a condition useful later on.

In all cases, it is easily seen that $	\bm{q}_{21}  \times \bm{q}_{41} = \bm{q}_{41}  \times \bm{q}_{31}$. Then, using $p_{21} = p_{31}$ and $p_{42} = p_{43}$, Eqs.(\ref{4cc:a}) and (\ref{4cc:b}) imply immediately that $m_2 = m_3$. This is because $p_{12} \neq p_{23}$ by the assumption that $\triangle 123 $ is not equilateral. This completes the proof of the proposition.

Next, we prove that the ratio of the two masses at the ends of the axis of symmetry is a function of the angles defining the specific shape of the kite. Each kite shape (up to scaling and rotation) defines a unique ratio of masses.

The symmetry of the kite and the equality $m_2 = m_3$ make Eq.(\ref{4cc:d}) redundant. Furthermore, Eq.(\ref{4cc:e}) and Eq.(\ref{4cc:f}) are identical, hence one of them is also redundant. We are left with two independent equations, e.g., Eq.(\ref{4cc:b}) and (\ref{4cc:f}). When rewritten in terms of $q_{21} (=q_{31})$ and $q_{42} (=q_{43})$), these become:
\begin{subequations} \label{isos}
	\begin{align}
			m_4 \left[ \dfrac{1}{q_{42}^3} - \dfrac{1}{q_{41}^3}\right]\left( \bm{q}_{41}  \times \bm{q}_{31}\right) &= m_2 \left[ \dfrac{1}{q_{21}^3} - \dfrac{1}{q_{32}^3}\right]\left( \bm{q}_{21}  \times \bm{q}_{31}\right)
		\label{isos:a}\\
		m_1 \left[ \dfrac{1}{q_{21}^3} - \dfrac{1}{q_{41}^3}\right]\left( \bm{q}_{41}  \times \bm{q}_{31}\right) &= m_2 \left[ \dfrac{1}{q_{42}^3} - \dfrac{1}{q_{32}^3}\right]\left( \bm{q}_{42}  \times \bm{q}_{32}\right)
		\label{isos:b}
	\end{align}
\end{subequations}
We must now distinguish between the convex and concave cases. 

\subsection{Convex kite}
\label{sec:convex}

\begin{figure}[h!]
	\label{fig:kitea}
	\includegraphics[width=0.7\linewidth]{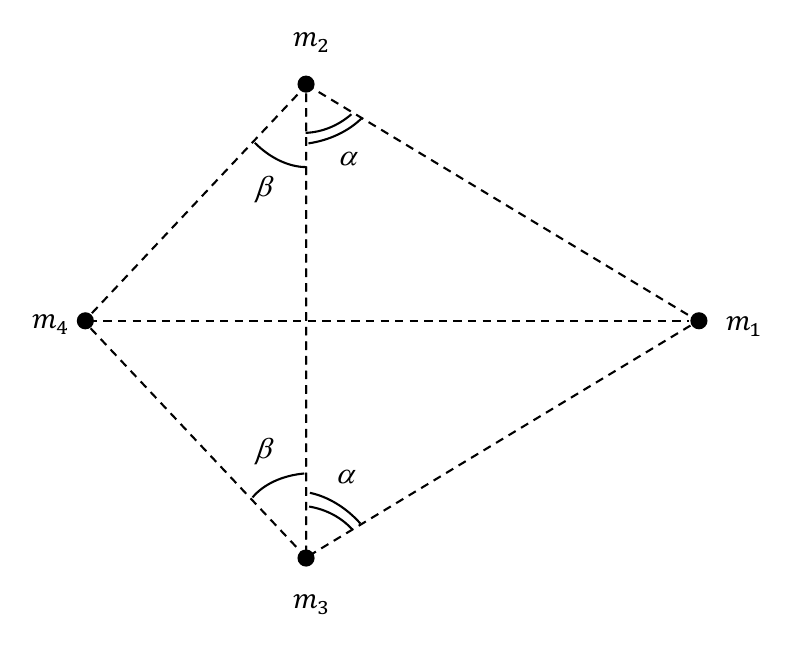}
	\caption{The first of three possible configurations of a kite. $m_4$ is outside the triangle $\triangle 123$, to its left, and the kite is convex. The angle $\alpha$, between $\bm{q}_{12}$ and $\bm{q}_{23}$ is marked with a double line, while the angle $\beta$, between $\bm{q}_{24}$ and $\bm{q}_{23}$, is marked with a single line.}
	\label{Fig:convex_kite}
\end{figure}

When $m_1$ and $m_4$ are on opposite sides of the vector $\bm{q}_{23}$, the kite is convex, as depicted in Fig.\ref{Fig:convex_kite}. We shall now obtain an algebraic relation between the ratios of the bodies' masses and the angles that define the shape of the kite.

Referring to Fig.\ref{Fig:convex_kite}, the angle between $\bm{q}_{12}$ and $\bm{q}_{14}$ is $\dfrac{\pi}{2} - \alpha$, and so is the angle between $\bm{q}_{13}$ and $\bm{q}_{14}$. The angle between $\bm{q}_{24}$ and $\bm{q}_{23}$ is $\beta$ and the angle between $\bm{q}_{12}$ and $\bm{q}_{13}$ is $\pi - 2 \alpha$.

All the vector products appearing in Eqs.(\ref{isos}) point in the same direction; taking the norms of the equations, we obtain
\begin{subequations} \label{kitevex}
	\begin{align}
		m_4 \left[ \dfrac{1}{q_{24}^3} - \dfrac{1}{q_{14}^3}\right]q_{41} q_{21} \cos \alpha &= m_2 \left[ \dfrac{1}{q_{21}^3} - \dfrac{1}{q_{32}^3}\right] \left( q_{12}\right)^2 \sin (2 \alpha)
		\label{kitevex:a}\\
		m_1 \left[ \dfrac{1}{q_{21}^3} - \dfrac{1}{q_{41}^3}\right]q_{41} q_{21} \cos \alpha &= m_2 \left[ \dfrac{1}{q_{42}^3} - \dfrac{1}{q_{32}^3}\right]q_{42} q_{32} \sin \beta
		\label{kitevex:b}
	\end{align}
\end{subequations}
Here we replaced $q_{13}$ by its equal $q_{12}$ and $q_{34}$ by $q_{24}$, to reduce the number of variables.

These equations feature a mixture of distances and angles, but can be expressed through angles alone, or through (ratios of) distances alone. The use of angles makes the scale invariance more explicit, and I shall therefore use them as fundamental variables. Pair-space leads us naturally to think of a configuration as a set of triplets, i.e., of triangles, in which the distances relate to the angles through the sine theorem. Thus, considering the triangles $\triangle 124$ and  $\triangle 123$ we obtain the relations:

\begin{subequations} \label{sinekite}
	\begin{align}
	q_{14} &= q_{12}\dfrac{\sin \left(\alpha + \beta \right)}{\cos \beta}   &\Rightarrow \qquad p_{14} &= p_{12}\dfrac{\cos^3 \beta}{\sin^3 \left(\alpha + \beta \right)} 
	\label{sinekite:a}\\
	q_{24} &= q_{12}\dfrac{\cos \alpha }{\cos \beta}  &\Rightarrow \qquad  p_{24} &= p_{12}\dfrac{\cos^3 \beta}{\cos^3 \alpha} 
	\label{sinekite:b}\\
	q_{23} &= 2 q_{12}\cos \alpha  &\Rightarrow \qquad   p_{23} &= p_{12}\dfrac{1}{8\cos^3 \alpha} 
	\label{sinekite:c}
	\end{align}
\end{subequations}

Substituting these into Eqs.(\ref{kitevex}), we obtain, after some trivial algebra, relations for the masses and angles:

\begin{subequations} \label{kitemass}
	\begin{align}
		\dfrac{m_1}{m_2} &=  \dfrac{\sin \beta \sin^2 \left( \alpha + \beta \right) \left[8	cos^3 \beta - 1\right]}{4 \cos^2  \alpha \left[\sin^3 \left(\alpha + \beta\right)- \cos^3 \beta \right]}
		\label{kitemass:a}\\
		\dfrac{m_4}{m_2} &=  \dfrac{\sin \alpha \sin^2 \left( \alpha + \beta \right) \left[8 cos^3 \alpha - 1\right]}{4 \cos^2 \beta \left[\sin^3 \left(\alpha + \beta\right)- \cos^3 \alpha \right]}
		\label{kitemass:b}
	\end{align}
\end{subequations}

These equations are symmetrical under the exchange $m_1 \longleftrightarrow m_4$ and $\alpha \longleftrightarrow \beta$, which represents a mirror reflection of the kite along the vertical line from $m_2$ to $m_3$ (or a spatial rotation around that line), and a consequent relabeling of vertices and angles.

These equations can be considered in two complementary ways. The first is the ``direct'' problem, which means solving for the angles of the kite given the masses of the bodies. The ``inverse'' problem is finding the masses given the shape of the kite, i.e., the angles $\alpha, \beta$. 

The direct problem is notoriously difficult, its main challenge being to prove that a given set of masses defines a unique kite configuration. Albouy resolved the problem for the case of four equal masses \citep{Albouy1, Albouy2}. Some of the important works discussing other mass distributions or the general case are \citet{leandro,Corbera,Santoprete,Sun et. al.,Roberts}. In the present work, the stress is on showcasing the value of the pair-space approach as a systematizing and organizing approach. Trying to tackle the direct problem would take us too far afield, due to the complexity of the question. Future work may discuss this question. Here, in order to keep the size of the paper manageable and its structure relatively readable, we shall concentrate instead on the easier inverse problem, i.e., given a kite shape, finding the masses that realize this geometry.

The requirement that the masses are positive limits the range of possible angles. We shall now examine the possible signs of the numerator and denominator of Eqs.(\ref{kitemass}), taking into account that a ratio of masses must be positive. 

Consider first the possibility that the numerator of $m_4/m_2$ is negative, i.e., $8 cos^3 \alpha - 1 < 0$, or $\alpha > \dfrac{\pi}{3}$. Then the denominator must also be negative, i.e., $\sin^3 \left(\alpha + \beta\right)- \cos^3 \alpha < 0$. This condition has two solutions, both of which are impossible: 

\begin{subequations}
\begin{align*}
	\alpha + \beta  &> \dfrac{\pi}{2} + \alpha  &\Rightarrow& \qquad \beta > \dfrac{\pi}{2}  \\
	\alpha + \beta  &< \dfrac{\pi}{2} - \alpha  &\Rightarrow& \qquad \beta + 2 \alpha < \dfrac{\pi}{2}  	
\end{align*}
\end{subequations}
The first case solution contradicts the condition that $ \beta < \dfrac{\pi}{2}$. The second contradicts the negativity of the numerator, which requires that $\alpha > \dfrac{\pi}{3}$. But this last condition already implies that $2 \alpha > \dfrac{\pi}{2}$ even before adding $\beta$. Therefore, we must have that $\alpha < \dfrac{\pi}{3}$ and the numerator and denominator in Eq.(\ref{kitemass:b}) are both positive. By the $m_1 \longleftrightarrow m_4, \alpha \longleftrightarrow \beta$ symmetry mentioned above, the same argument shows that also $\beta < \dfrac{\pi}{3}$.

The mass ratio becomes singular if $\sin^3 \left(\alpha + \beta\right)- \cos^3\alpha = \sin^3 \left(\alpha + \beta\right)- \cos^3\beta = 0$. The solution to this condition is $\alpha = \beta = \dfrac{\pi}{6}$ and $m_2 \rightarrow 0$. Since we do not allow zero masses, this solution must be discarded (this will happen automatically, however, see below).

The remaining conditions are thus that both numerators and denominators in Eqs.(\ref{kitemass}) are positive. This implies that 

\begin{subequations}
	\begin{align*}
		8 cos^3 \beta - 1 &> 0  &\Rightarrow& \qquad \beta < \dfrac{\pi}{3} \\
		\dfrac{\pi}{2} + \beta > \alpha + \beta  &> \dfrac{\pi}{2} - \beta  &\Rightarrow& \qquad \dfrac{\pi}{2} > \alpha > \dfrac{\pi}{2} - 2 \beta  \\		
		8 cos^3 \alpha - 1 &> 0  &\Rightarrow& \qquad \alpha < \dfrac{\pi}{3} \\
		\dfrac{\pi}{2} + \alpha > \alpha + \beta  &> \dfrac{\pi}{2} - \alpha  &\Rightarrow& \qquad \dfrac{\pi}{2} > \beta > \dfrac{\pi}{2} - 2 \alpha    	
	\end{align*}
\end{subequations}

\begin{figure}[h!]
	\includegraphics[width=0.7\linewidth]{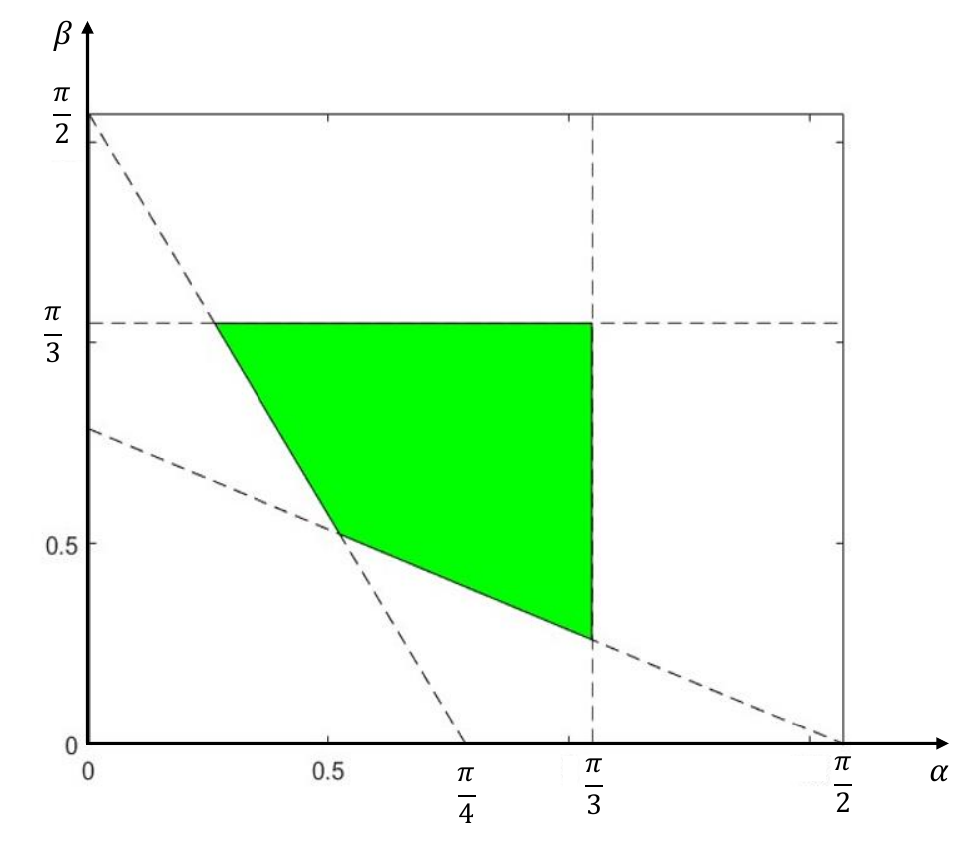}
	\caption{The areas of possible angles for a central kite configuration in the case of a convex kite. The allowed areas are shaded in the graph. The lines bordering them are outside the allowed range as they correspond to singular conditions, such as some masses vanishing or two masses sitting one on top of the other. We explicitly exclude such configurations here.}
	\label{Fig:kite_convex_angles}
\end{figure}

The possible solutions for the angles are depicted in Fig.(\ref{Fig:kite_convex_angles}). The possible angles occupy an area bordered by the vertical line $\alpha = \dfrac{\pi}{3}$, the horizontal line $\beta = \dfrac{\pi}{3}$, and the two straight lines $\alpha = \dfrac{\pi}{2} - 2 \beta$ and $\alpha = \dfrac{\pi}{4} - \dfrac{\beta}{2}$. The borders themselves are not included in the area, as they correspond to singular cases (infinitely long kite, zero masses and the like). In particular, the singular case $\alpha = \beta = \dfrac{\pi}{6}$ mentioned above (corresponding to $m_2 = 0$) is the meeting point of the two diagonal lines and is therefore excluded from possible solutions automatically. Note that in \citet{Erdi and Czirjak}, the area of possible solutions is half of what is depicted here, because they assumes that $\alpha \geq \beta$.

\subsection{Rhombus}
\label{subsection:rhombus}

\begin{figure}[h!]
	\includegraphics[width=0.7\linewidth]{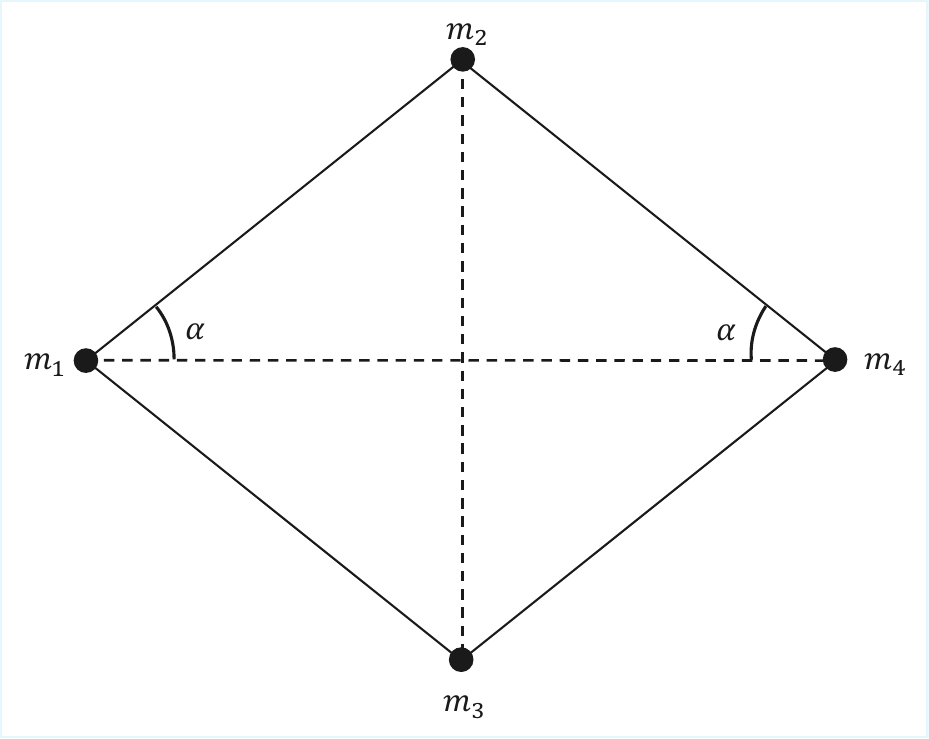}
	\caption{The rhombus configuration. Up to scaling, the figure is entirely determined by the angle $\alpha$.}
	\label{Fig:Rhombus}
\end{figure}

The case $\alpha = \beta > \dfrac{\pi}{6}$ corresponds to a rhombus (see Fig.\ref{Fig:Rhombus}). Eq.(\ref{sinekite:b}) shows that indeed $q_{24} = q_{12}$, so that all the edges are equal. Substituting $\alpha = \beta$ in Eqs.(\ref{kitemass}) shows that the two equations become identical, hence $m_1 = m_4$. Therefore, in a rhombus, there are two pairs of equal masses, located opposite one another. The single ratio $\dfrac{m_1}{m_2}$ determines the angle $\alpha$ and vice versa. Eq.(\ref{kitemass:a}), for example, becomes

\begin{equation}
	T(\alpha) = \dfrac{sin^3 \alpha \left(8 \cos^3 \alpha - 1 \right)} {cos^3 \alpha\left(8 \sin^3 \alpha - 1 \right)} = \dfrac{1-\dfrac{1}{8 \cos^3 \alpha}}{1-\dfrac{1}{8 \sin^3 \alpha}} = \dfrac{m_1}{m_2}
\end{equation}

As expected, $T(\alpha)$ is positive only in the range $\dfrac{\pi}{6} < \alpha < \dfrac{\pi}{3}$. In this range, $\cos(\alpha)$ is monotonically decreasing, hence $(2\cos(\alpha))^{-3}$ is monotonically increasing; therefore, the numerator of $T(\alpha)$, $1-\dfrac{1}{8 \cos^3 \alpha}$ is monotonically decreasing. A similar argument shows that the denominator, $1-\dfrac{1}{8 \sin^3 \alpha}$ is monotonically increasing in this range, since $\sin(\alpha)$ is monotonically increasing itself. Hence, both numerator and denominator contribute to make $T(\alpha)$ a monotonically decreasing function, which means that every ratio $m_1/m_2$ corresponds to a unique angle $\alpha$ and vice versa. Up to scaling, this determines completely the shape of the rhombus. Note that for this particular shape, the equations are simple enough to solve the direct problem. Every mass ratio determines a single angle $\alpha$, thus for every set of masses for which the rhombus exists, it is unique.

\subsection{Concave kite}
\label{sec:concavekite}
If $m_1$ and $m_4$ are on the same side of the vector $\bm{q}_{23}$, the kite is concave. The angles $\alpha$ and $\beta$ are defined as before. Here as well, we seek a relation between the mass ratios and the angles defining the kite's shape. 

\begin{figure}[h!]
	\includegraphics[width=0.7\linewidth]{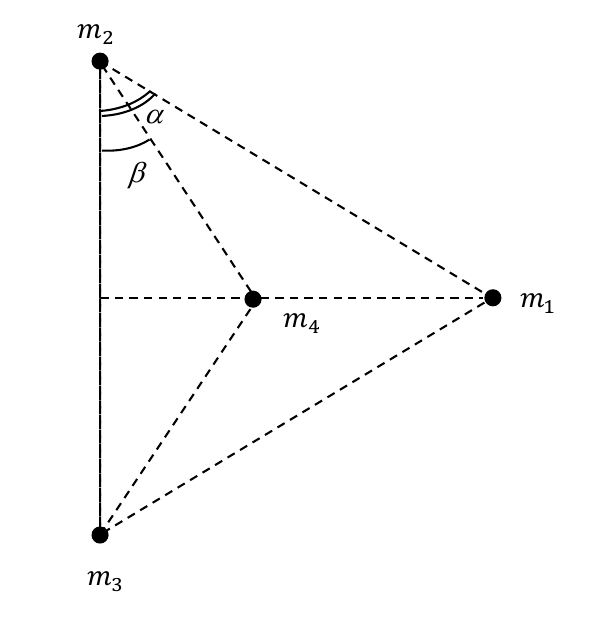}
	\caption{The second possible kite configuration. $m_4$ and $m_1$ are on the same side of $\bm{q}_{12}$ and $m_4$ is inside the triangle. The resulting kite is concave. As before, the angle $\alpha$, between $\bm{q}_{12}$ and $\bm{q}_{23}$, is marked with a double line, while the angle $\beta$, between $\bm{q}_{24}$ and $\bm{q}_{23}$, is marked with a single line. Note that in this case $\beta < \alpha$. }
	\label{Fig:concave1}
\end{figure}

\begin{figure}[h!]
	\includegraphics[width=0.7\linewidth]{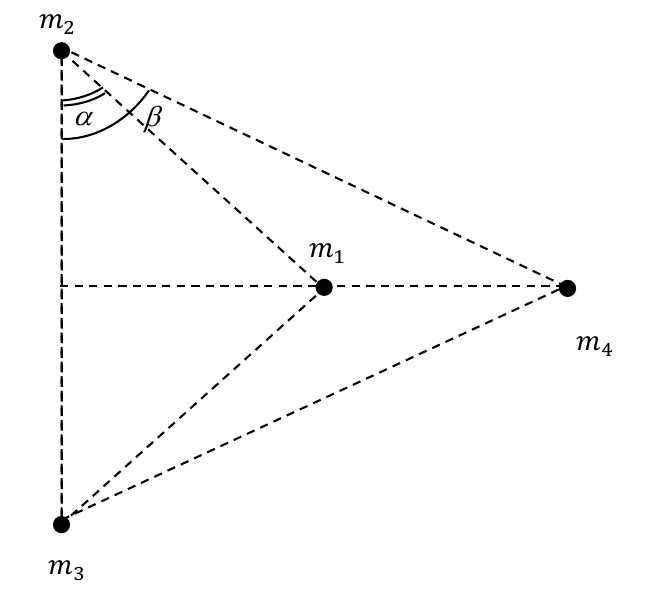}
	\caption{The third possible kite configuration, where $m_4$ and $m_1$ are on the same side of $\bm{q}_{12}$, but $m_4$ is outside the triangle. Once again, the angle $\alpha$, between $\bm{q}_{12}$ and $\bm{q}_{23}$, is marked with a double line, while the angle $\beta$, between $\bm{q}_{24}$ and $\bm{q}_{23}$, is marked with a single line. Here, $\alpha < \beta$, but note that this case is identical to the case in Fig.(\ref{Fig:concave1}) if we relabel $m_1$ as $m_4$ and vice versa, and $\alpha$ as $\beta$ and vice versa.}
	\label{Fig:concave2}
\end{figure}

For any set of masses there are in principle two possible concave kites. In Fig.\ref{Fig:concave1}, $m_4$ is to the left of $m_1$, inside the triangle $\triangle 123$, and $\beta < \alpha$. In Fig.\ref{Fig:concave2}, $m_4$ is to the right of $m_1$, outside the triangle $\triangle 123$, and  $\beta > \alpha$. As long as we consider the masses as arbitrary parameters, these two cases are equivalent under an exchange of labels $1 \longleftrightarrow 4$, and $\beta \longleftrightarrow \alpha$. hence we need not distinguish between them, and in the following I refer to Fig.\ref{Fig:concave1}, i.e., I assume that $\alpha > \beta$.

Taking the norm of Eqs.(\ref{isos}), and considering the directions of the cross products, we have 
\begin{subequations} \label{kitecav}
	\begin{align}
		m_4 \left[ p_{24} - p_{14} \right]q_{41} q_{21} \cos \alpha &= m_2 \left[ p_{21} - p_{32} \right] \left( q_{12}\right)^2 \sin (2 \alpha)
		\label{kitecav:a}\\
		- m_1 \left[ p_{21} - p_{41} \right]q_{41} q_{21} \cos \alpha &= m_2 \left[ p_{42} - p_{32} \right]q_{42} q_{32} \sin \beta
		\label{kitecav:b}
	\end{align}
\end{subequations}
Notice the negative sign in Eq.(\ref{kitecav:b}) due to the opposite directions of the vector products that appear in it. As before, we replaced $q_{13} \Longrightarrow q_{12}$ and $q_{34} \Longrightarrow q_{24}$.

Applying the sine theorem to $\triangle 124$ and  $\triangle 123$ we obtain the relations:

\begin{subequations} \label{sinekcav}
	\begin{align}
		q_{14} &= q_{12}\dfrac{\sin \left(\alpha - \beta \right)}{\cos \beta}   &\Rightarrow \qquad p_{14} &= p_{12}\dfrac{\cos^3 \beta}{\sin^3 \left(\alpha - \beta \right)} 
		\label{sinekcav:a}\\
		q_{24} &= q_{12}\dfrac{\cos \alpha }{\cos \beta}  &\Rightarrow \qquad  p_{24} &= p_{12}\dfrac{\cos^3 \beta}{\cos^3 \alpha} 
		\label{sinekcav:b}\\
		q_{23} &= 2 q_{12}\cos \alpha  &\Rightarrow \qquad   p_{23} &= p_{12}\dfrac{1}{8\cos^3 \alpha} 
		\label{sinekcav:c}
	\end{align}
\end{subequations}

As in the convex case, substituting these into Eqs.(\ref{kitecav}) yields relations for the masses and angles:

\begin{subequations} \label{kitecavmass}
	\begin{align}
		\dfrac{m_1}{m_2} &=  \dfrac{\sin \beta \sin^2 \left( \alpha - \beta \right) \left[1 - 8	cos^3 \beta \right]}{4 \cos^2  \alpha \left[\sin^3 \left(\alpha - \beta \right) - \cos^3 \beta \right]}
		\label{kitecavmass:a}\\
		\dfrac{m_4}{m_2} &=  \dfrac{\sin \alpha \sin^2 \left( \alpha - \beta \right) \left[8 cos^3 \alpha - 1\right]}{4 \cos^2 \beta \left[\sin^3 \left(\alpha - \beta\right)- \cos^3 \alpha \right]}
		\label{kitecavmass:b}
	\end{align}
\end{subequations}

As in the convex case, the direct problem, i.e., determining $\alpha$ and $\beta$ from the mass ratios is very difficult. In particular, one would like to know whether the cases described in Figs.(\ref{Fig:concave1}) and (\ref{Fig:concave2}) are both possible for a given set of masses and fixed labeling, under what conditions the solution is unique and so on. Some of the notable investigations of this problem are \citet{Bernat,Alvarez-Ramirez and Llibre,Deng and Zhang,Kovari and Erdi,Liu}. In the present work, we shall only treat the inverse problem, as we did in the convex case and for the same reasons. 

Once again, the requirement that the masses are positive limits the range of possible angles. First, since $\alpha < \dfrac{\pi}{2}$, we have that $\alpha - \beta < \dfrac{\pi}{2} - \beta$, and therefore $\sin \left( \alpha - \beta \right) < \cos \beta$. Hence, the denominator in Eq.(\ref{kitecavmass:a}) is negative. The numerator must then also be negative, so that 
\begin{equation}
1 - 8 \cos^3 \beta < 0  \qquad  \Longrightarrow \qquad \beta < \dfrac{\pi}{3} 
\end{equation}

In Eq.(\ref{kitecavmass:b}), the ratio on the right hand side must be positive, and this can happen in two ways.

\textbf{Case 1}: Numerator and denominator both positive. Then
\begin{equation*}
8 \cos^3 \alpha - 1 > 0  \qquad  \Longrightarrow \qquad \alpha < \dfrac{\pi}{3} 
\end{equation*}
and
\begin{equation*}
\sin \left(\alpha - \beta\right) > \cos \alpha \qquad \Longrightarrow  \qquad \dfrac{\pi}{2} + \alpha > \alpha - \beta > \dfrac{\pi}{2} - \alpha
\end{equation*}
The condition $ \dfrac{\pi}{2} + \alpha > \alpha - \beta $ is automatic since $\beta > 0$. Thus, we have that
\begin{equation*}
	\text{case 1 =}
	\begin{cases}
		\alpha < \dfrac{\pi}{3} \\
		2 \alpha - \beta > \dfrac{\pi}{2}
	\end{cases}
\end{equation*}

\textbf{Case 2}: Numerator and denominator both negative. Then
\begin{equation*}
	8 \cos^3 \alpha - 1 < 0  \qquad  \Longrightarrow \qquad \alpha > \dfrac{\pi}{3} 
\end{equation*}
and
\begin{equation*}
	\sin \left(\alpha - \beta\right) < \cos \alpha \qquad \Longrightarrow  \qquad \dfrac{\pi}{2} + \alpha < \alpha - \beta  \qquad \text{or}  \qquad \alpha - \beta < \dfrac{\pi}{2} - \alpha
\end{equation*}
The case $\dfrac{\pi}{2} + \alpha < \alpha - \beta$ holds only if $- \beta > \dfrac{\pi}{2}$, which is impossible. Hence,
\begin{equation*}
		\text{case 2 =}
	\begin{cases}
		\alpha > \dfrac{\pi}{3} \\
		2 \alpha - \beta < \dfrac{\pi}{2}
	\end{cases}
\end{equation*}

\begin{figure}[h!]
		\includegraphics[width=0.7\linewidth]{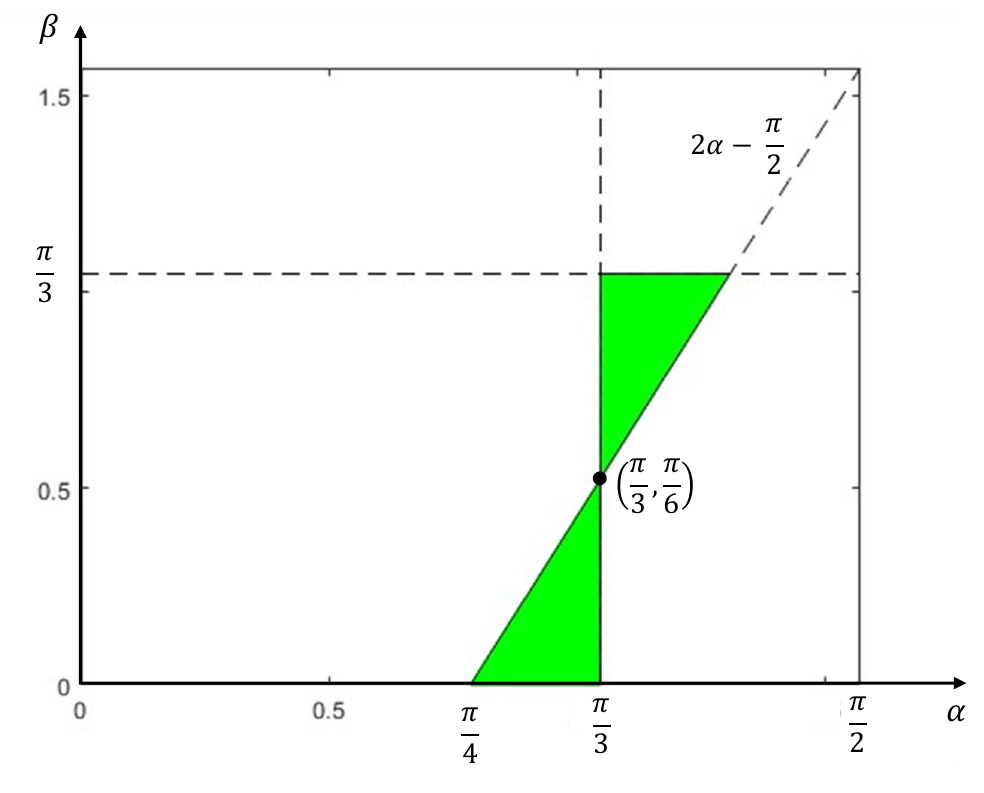}
		\caption{The areas of possible angles for a concave kite configuration. The allowed areas are shaded in the graph. The lines bordering them are outside the allowed range as they correspond to singular conditions, such as some masses vanishing or two masses sitting one on top of the other. We explicitly exclude such configurations here.}
		\label{Fig:kite_concave}
	\end{figure}

These two cases represent two triangular areas in a $\beta$ vs $\alpha$ plot, as described in Fig.~\ref{Fig:kite_concave}. In the range $\dfrac{\pi}{4} < \alpha < \dfrac{\pi}{3}$, the possible region lies below the straight line  $\beta = 2 \alpha - \dfrac{\pi}{2}$; in the range $\dfrac{\pi}{3} < \alpha < \dfrac{5\pi}{12}$, the possible region lies above that same line, and below the horizontal limit $\beta = \dfrac{\pi}{3}$.

The intersection point between the lines $\alpha = \dfrac{\pi}{3}$ and $\beta = 2 \alpha - \dfrac{\pi}{2}$ does not belong to the present case since it implies that $\dfrac{m_4}{m_2} = 0$, by Eq.(\ref{kitecavmass:b}). Nevertheless, this point does correspond geometrically to a central configuration, since $\alpha = \dfrac{\pi}{3}$ implies that the triangle $\triangle 123$ is equilateral. Eq.(\ref{kitecavmass:a}) holds and yields that $m_1 = m_2$, as it should since all the masses in the equilateral triangle must be equal. The condition $\beta = 2\alpha - \dfrac{\pi}{2} = \dfrac{\pi}{6}$ then places $m_4$ in the center of the triangle $\triangle 123$. We know from section \ref{sec:equilateral} that $m_4$ is then arbitrary, which accords with the failure of Eq.(\ref{kitecavmass:b}).
	
These results agree with the analysis of \citet{Erdi and Czirjak}, though their analysis relied on a different approach. \'{E}rdi and Czirj\'{a}k distinguish between two cases, depending on whether the center of mass of the system lies between $m_1$ and $m_4$ or not. Each case generates a different equation and these must be treated separately. The two triangular areas represent the solutions to these two different equations. In the present derivation, there is no need to distinguish these two cases, as they are governed by a single equation. The two regions emerge as two distinct solution branches to that unique equation.

\section{Parallelogram and isosceles trapezium}
\label{sec:parallel}

We have examined the possible configurations if there exists in the system a triangle that is at least isosceles. We now assume that it is not the case. Nevertheless, it may still be that two mutual distances are equal, provided that they do not both belong to any single triangle. Without loss of generality, we can label the two ends of one such distance $m_1$ and $m_2$. Then, since no triangle may be isosceles, the other distance may not share any vertex with the first. Therefore, it must be the distance between $m_3$ and $m_4$. In other words, without loss of generality, we can always assume that $q_{12} = q_{34}$.

Let us examine the possibility that there exists another pair of equal distances. Then once again, the sets of their vertices must be disjoint. One of the distances must extend between $m_1$ and some other mass, therefore. $m_2$ is already taken, so there are two possibilities. Either $q_{14} = q_{23}$ or $q_{13} = q_{24}$. We shall show that the first option corresponds to the already investigated rhombus shape, while the second corresponds to an isosceles trapezium. Let us now consider the possible arrangements of the masses.

To begin, we shall prove the following lemma:

\textbf{Lemma:} A central configuration with $q_{12}=q_{34}$ and either $q_{13} = q_{24}$ or $q_{14} = q_{23}$ must be convex.

\textbf{Proof:}

\begin{figure}[h!]
	\includegraphics[width=0.7\linewidth]{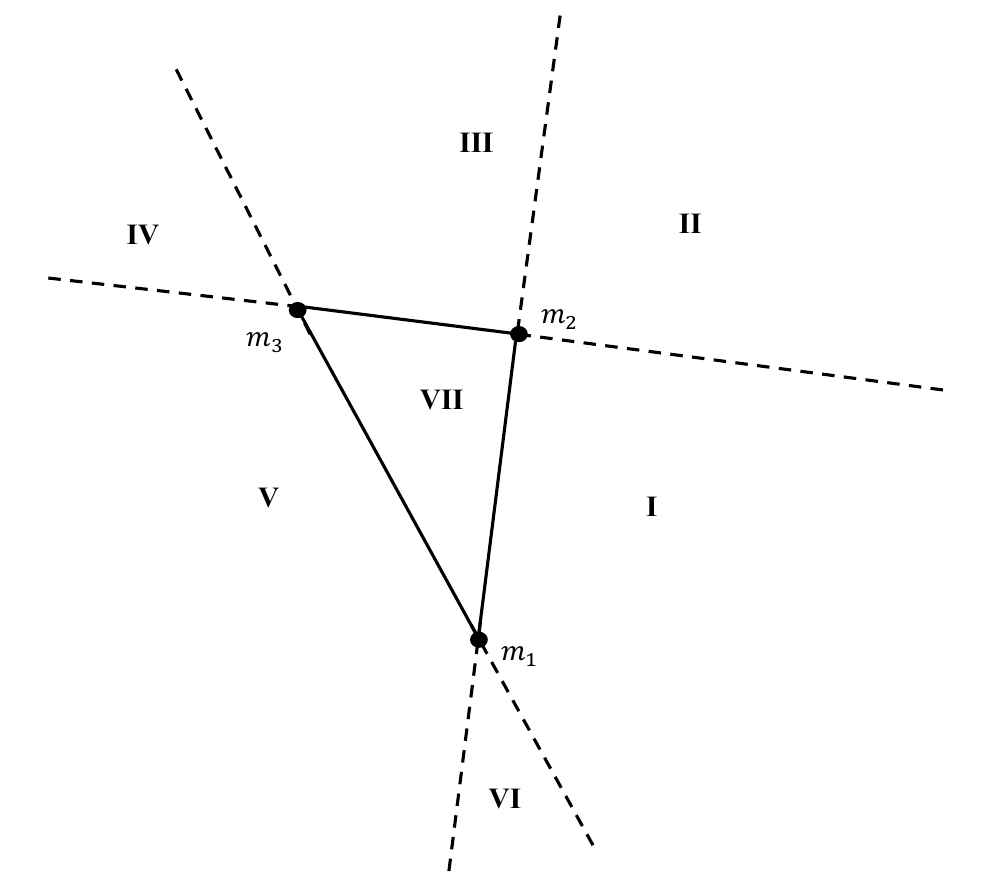}
	\caption{The possible areas in which to place a mass $m_4$, given a general triangle $\triangle 123$. Areas I-VI are outside the triangle. Area VII is inside it. Some positions of $m_4$ generate concave configurations, others generate convex configurations.}
	\label{Fig:paratrap}
\end{figure}

We start, arbitrarily, with the triangle $\triangle 123$. Extending the edges of the triangle to infinity divides the plane into seven areas, numbered as in Fig.\ref{Fig:paratrap}. The first six lie outside the triangle, and area VII is  inside. These are the possible locations of the fourth mass, $m_4$. 

\begin{figure}[h!]
	\includegraphics[width=0.7\linewidth]{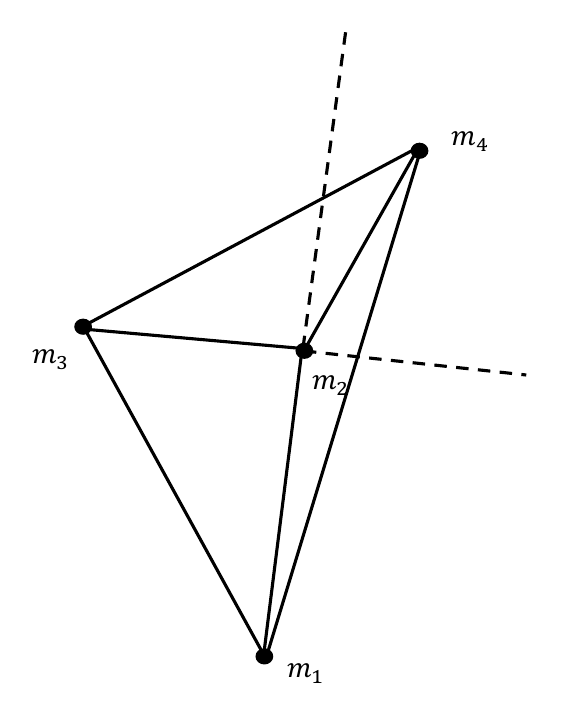}
	\caption{An attempt to place $m_4$ in area II of Fig.\ref{Fig:paratrap}. Here, $q_{12} = q_{34}$. The desired condition $q_{13}=q_{24}$ is violated, however. This is an instance of the impossibility of creating concave configurations with the appropriate requirements.}
	\label{Fig:concavepara}
\end{figure}

For a configuration to be convex, $m_4$ must only lie in areas I, III or V. Assume then that the configuration is concave and that $m_4$ is placed in zone II. The resulting shape is depicted in fig.\ref{Fig:concavepara}. We assume that $q_{12} = q_{34}$ and consider first the option that $q_{13} = q_{24}$. Then, referring to fig.\ref{Fig:concavepara} we have that $\triangle 123$ and $\triangle 234$ are SSS congruent since they share a common side $q_{23}$. This is clearly not the case in Fig.\ref{Fig:concavepara}, which suggests that this situation is impossible. Indeed, congruence implies that $\triangle 234$ has the same area as $\triangle 123$. However, $\triangle 234$ is strictly contained within $\triangle 123$, making its area strictly smaller. Therefore we have a contradiction and this shape cannot exist.

Next, we consider the second option, $q_{14} = q_{23}$. This time, $\triangle 124 \cong \triangle 234$. Denoting

\begin{subequations}
	\begin{align*}
		\measuredangle 243 &= \measuredangle 124 := \theta ,  \\
		\measuredangle 214 &= \measuredangle 234 := \psi    , 
	\end{align*}
\end{subequations}
we see that in $\triangle 124$, $\measuredangle 142 = \pi - (\theta + \psi)$. Therefore, $\measuredangle 143 = 2\pi - \theta - (\pi - \theta - \psi) = \pi  +\psi$. But then, triangle $\triangle 134$ has an interior angle larger than $\pi$, which is impossible. Since we found that both shapes cannot exist, there is no way to place $m_4$ in area II so that two pairs of distances are equal. 

This argument eliminates all the other concave possibilities, because they are identical to the shape we just considered under an appropriate relabeling of the masses. For example, it is easily seen that placing $m_4$ in area VII produces an identical shape if we relabel $m_1 \longleftrightarrow m_3$ and $m_2 \longleftrightarrow m_4$. This is because under such relabeling, the condition $q_{12} = q_{34}$ remains invariant, and so does $q_{13} = q_{24}$. Thus, the conditions on the shape are identical to the case just considered and its impossibility goes through in the same way. All other concave cases are eliminated by analogous arguments. This proves the lemma. 

Only convex options remain, therefore. Here again, one case suffices. Placing $m_4$ in area I, III or V produces shape that only differ by the labeling of the masses. Without loss of generality, we can choose to place $m_4$ in area V, therefore.

\subsection{Parallelogram}

\begin{figure}[h!]
	\includegraphics[width=0.7\linewidth]{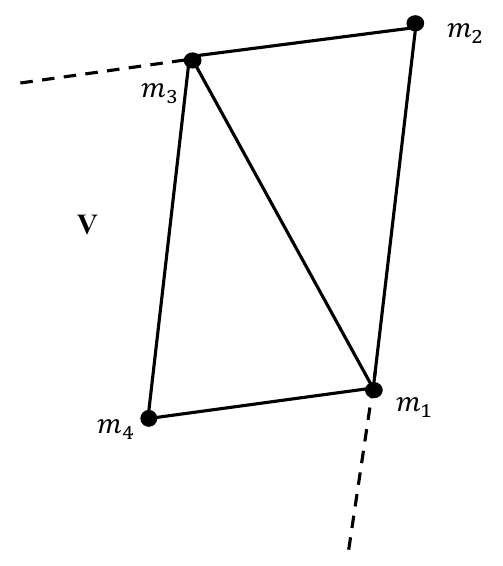}
	\caption{Placing $m_4$ in area V of Fig.\ref{Fig:paratrap} generates a parallelogram.}
	\label{Fig:parallel}
\end{figure}

Consider first the case $q_{12} = q_{34}$ and $q_{14} = q_{23}$. The situation is depicted in Fig.\ref{Fig:parallel} and we immediately see that $\square 1234$ is a parallelogram. We shall now prove that the only possible parallelogram is a rhombus.

Substituting $q_{12} = q_{34}$ and $q_{14} = q_{23}$ into Eq.(\ref{4cc:b}) yields

\begin{equation}
	\label{paralcase1}
\left( \dfrac{1}{q_{21}^3} - \dfrac{1}{q_{32}^3}\right) \left[m_2 \left( \bm{q}_{21}  \times \bm{q}_{31}\right) - m_4 \left( \bm{q}_{41}  \times \bm{q}_{31}\right) \right] = 0           
\end{equation}
The first possibility is that 
\begin{equation}
	\label{paral01}
m_2 \left( \bm{q}_{21}  \times \bm{q}_{31}\right) - m_4 \left( \bm{q}_{41}  \times \bm{q}_{31}\right) = 0 
\end{equation} 
A look at Fig. \ref{Fig:parallel} shows that $\bm{q} _{41}  \times \bm{q}_{31}$ is directed into the depicted surface, but $\bm{q} _{21}  \times \bm{q}_{31}$ points in the opposite direction. Therefore, Eq.(\ref{paral01}) cannot hold.

We must then have the other option, namely, that 
 \begin{equation}
 	\dfrac{1}{q_{21}^3} - \dfrac{1}{q_{32}^3} = 0  \qquad ,
 \end{equation}
and hence, that $q_{21} = q_{32}$. This makes the parallelogram into a rhombus. This case has been treated already in section \ref{subsection:rhombus}. Thus we conclude that the rhombus is the only central configuration shaped as a parallelogram. It requires two pairs of equal masses, each member of the pair facing the other at the opposite vertex. 

\subsection{Isosceles trapezium}

\begin{figure}[h!]
	\includegraphics[width=0.7\linewidth]{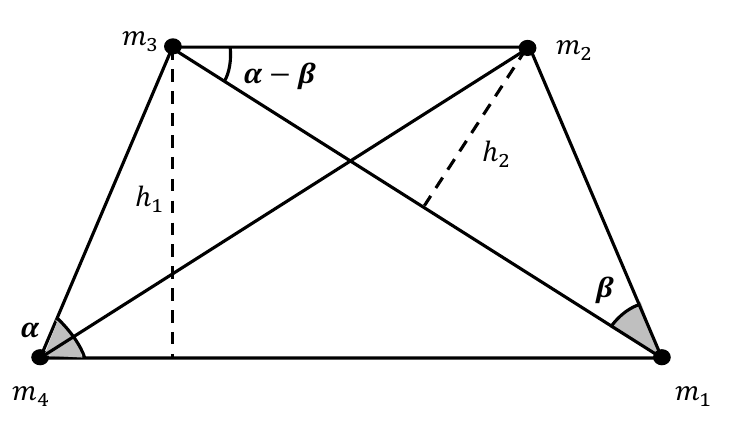}
	\caption{The isosceles trapezium configuration. Up to scaling, the shape is determined by two angles, shaded in the graph. $\alpha$ is the base angle, and $\beta$ the angle between the diagonal and the side.}
	\label{Fig:trapezium}
\end{figure}

We now consider the second case, namely $q_{12} = q_{34}$ and $q_{13} = q_{24}$. The situation is depicted in Fig.\ref{Fig:trapezium}. We see that $\square 1234$ is a quadrilateral with two equal opposing sides, and two equal diagonals. Therefore, it is an isosceles trapezium. We shall now prove:

\textbf{Proposition 5:}

1. The masses at the opposite ends of the trapezium's long base are equal.

2. The masses at the opposite ends of the trapezium's short base are equal.

\textbf{Proof:}

Up to scaling, an isosceles trapezium is defined by two angles. We choose the following (see Fig.\ref{Fig:trapezium}):
\begin{subequations}
	\begin{align}
		\measuredangle 214 = \measuredangle 341 := \alpha \\
		\measuredangle 213 = \measuredangle 342 := \beta 
	\end{align}
\end{subequations}

Referring to Fig. \ref{Fig:trapezium}, we consider the height of the trapezium, $h_1$ and the height $h_2$ of the triangle $\triangle 123$, from the vertex $m_2$ to the base $q_{13}$. We have that 
\begin{subequations}  \label{height}
	\begin{align}
		h_1 = q _{21} \sin \alpha = q_{31}\sin \left(\alpha - \beta\right) \label{height:a}  \\
		h_2 = q _{21} \sin \beta = q_{32}\sin \left(\alpha - \beta\right)  \label{height:b}
	\end{align}
\end{subequations}

Now take the norm of Eq.(\ref{4cc:c}). Taking into account the opposite directions of $\bm{q} _{21}  \times \bm{q}_{31}$ and $\bm{q}_{42}  \times \bm{q}_{32}$, we obtain that

\begin{equation}
	m_1 \left[ \dfrac{1}{q_{21}^3} - \dfrac{1}{q_{31}^3}\right]q _{21} q_{31}\sin \beta = m_4 \left[ \dfrac{1}{q_{43}^3} - \dfrac{1}{q_{42}^3}\right]q_{42}q_{32}\sin \left(\alpha - \beta\right) 
\end{equation}

First note that we cannot have $q_{21} = q_{31}$ since then $\triangle 123$ would be isosceles and we already showed that in this case the configuration must be a kite. Furthermore, since $q_{12} = q_{34}$ and $q_{13} = q_{24}$, we have that 
\begin{equation}
\dfrac{1}{q_{21}^3} - \dfrac{1}{q_{31}^3} = \dfrac{1}{q_{43}^3} - \dfrac{1}{q_{42}^3} \neq 0
\end{equation}
Canceling these expressions from the two sides of the equation and using again that $q_{13} = q_{24}$ leaves us with
\begin{equation}
	m_1 q _{21} \sin \beta = m_4 q_{32}\sin \left(\alpha - \beta\right)
\end{equation}
Using Eq.(\ref{height:b}), we are left with $m_1 = m_4$, i.e., the masses at the vertices of the long base of the trapezium must be equal.

To show the equivalent property for the short base, we repeat the argument for Eq.(\ref{4cc:d}), whose norm (again taking into account the opposite directions of the vector products) is:
\begin{equation}
		m_2 \left[ \dfrac{1}{q_{42}^3} - \dfrac{1}{q_{21}^3}\right]q _{21}  q_{41}\sin \alpha = m_3 \left[ \dfrac{1}{q_{31}^3} - \dfrac{1}{q_{43}^3}\right]q _{41}q_{31}\sin \left(\alpha - \beta \right)
\end{equation}

Once again eliminating equal terms on both sides leaves us with 
\begin{equation}
	m_2 q _{21} \sin \alpha = m_3 q_{31}\sin \left(\alpha - \beta\right)
\end{equation}
From Eq.(\ref{height:a}), we find $m_2 = m_3$. Hence the masses at the vertices of the short base of the trapezium must also be equal. This proves the proposition.

The equalities of the masses make some of Eqs.(\ref{4cc}) redundant. We shall now eliminate them. We compare Eqs.(\ref{4cc:a}) and (\ref{4cc:f}). Taking into account the different directions of the vector products, the norms of these equations become
\begin{subequations}   
	\begin{align}
	m_3 \left[ \dfrac{1}{q_{31}^3} - \dfrac{1}{q_{32}^3}\right] q_{31} q _{21} \sin \beta &= m_4 \left[ \dfrac{1}{q_{42}^3} - \dfrac{1}{q_{41}^3}\right]q_{41} q_{21}  \sin \alpha    . \\
	m_2 \left[ \dfrac{1}{q_{42}^3} - \dfrac{1}{q_{32}^3}\right]q_{42} q_{32} \sin \left(\alpha - \beta\right) & = m_1 \left[ \dfrac{1}{q_{31}^3} - \dfrac{1}{q_{41}^3}\right]q_{41} q_{31} \sin \left(\alpha - \beta\right)  .
	\end{align}
\end{subequations}
Using Eqs.(\ref{height}) and the previous result that $m_2 = m_3$ and $m_1 = m_4$, we can also write

\begin{subequations}   
	\begin{align}
		m_2 \left[ \dfrac{1}{q_{31}^3} - \dfrac{1}{q_{32}^3}\right] q_{31} h_2 &= m_1 \left[ \dfrac{1}{q_{42}^3} - \dfrac{1}{q_{41}^3}\right]q_{41} h_1       \\
		m_2 \left[ \dfrac{1}{q_{42}^3} - \dfrac{1}{q_{32}^3}\right]q_{42} h_2 & = m_1 \left[ \dfrac{1}{q_{31}^3} - \dfrac{1}{q_{41}^3}\right]q_{41} h_1  
	\end{align}
\end{subequations}
Since $q_{13} = q_{24}$, we see that both left hand sides of these equations are identical, and so are both right hand sides. Therefore, these equations are the same, and one of them is redundant. The exact same procedure shows that Eq.(\ref{4cc:b}) and (\ref{4cc:e}) are also identical.

We are thus left with two independent equations, which can be taken to be
\begin{subequations}  \label{m1m2} 
	\begin{align}
		m_2 \left[ \dfrac{1}{q_{43}^3} - \dfrac{1}{q_{32}^3}\right] q_{42} h_2 &= m_1 \left[ \dfrac{1}{q_{41}^3} - \dfrac{1}{q_{21}^3}\right]q_{41} h_1   ,    \label{m1m2:a}   \\
		m_2 \left[ \dfrac{1}{q_{42}^3} - \dfrac{1}{q_{32}^3}\right]q_{42} h_2 & = m_1 \left[ \dfrac{1}{q_{31}^3} - \dfrac{1}{q_{41}^3}\right]q_{41} h_1  .  \label{m1m2:b}
	\end{align}
\end{subequations}

From these equations, we now find the relation between the angles $\alpha, \beta$ and the ratio of masses. Dividing the two equations by each other yields a mass-independent relation. After a little rearranging, and recalling that $p_{ij} = q^{-3}_{ij}$, it can be written as
\begin{equation}
	\label{massless}
	\dfrac{p_{31} - p_{41}}{p_{42} - p_{32}} = \dfrac{p_{41}-p_{21}}{p_{43}-p_{23}}
\end{equation}

This result could have been obtained from one of the mass-independent general relations. If in Eq.(\ref{dziobek:d}) we replace on the right hand side $p_{34}$ by $p_{12}$ and $p_{13}$ by $p_{24}$, then the term $p_{24} - p_{12}$ cancels out and we are left with 
\begin{equation}
	- (p_{14}-p_{13})(p_{34}-p_{23}) =(p_{14}-p_{12})(p_{24}-p_{23}) .
\end{equation}
which is equivalent to Eq.(\ref{massless}).

To transform this relation into an equation for the angles $\alpha, \beta$, we look at the triangles $\triangle 123$ and $\triangle 124$. From the sine theorem in these, we obtain the following relations:

\begin{subequations}
	\label{sinetrap}
	\begin{align}
		q_{14} &= q_{12}\dfrac{\sin \left(2 \alpha - \beta\right)}{\sin \left(\alpha - \beta\right)}  \label{sinetrap:a}  \\
		q_{24} &= q_{13} = q_{12}\dfrac{\sin \alpha}{\sin \left(\alpha - \beta\right)}  \label{sinetrap:b}  \\
		q_{23} &= q_{12}\dfrac{\sin \beta}{\sin \left(\alpha - \beta\right)}  \label{sinetrap:c}
	\end{align}
\end{subequations}
Together with $q_{34} = q_{12}$, we can use these relations to rewrite Eq.(\ref{massless}) in terms of $\alpha, \beta$ only:

\begin{equation}
	\label{anglestrap}
	\dfrac{\sin^3 \alpha - \sin^3 \left(2 \alpha - \beta \right)}{\sin^3 \alpha - \sin^3 \beta} = \dfrac{\sin^3 \left(2 \alpha - \beta \right) - \sin^3 \left(\alpha - \beta \right)}{\sin^3 \left(\alpha - \beta \right) - \sin^3 \beta}   .
\end{equation}
Similarly, either one of Eqs.(\ref{m1m2}) will yield a second relation:
\begin{equation}
	\label{massratiotrap}
\dfrac{m_2}{m_1} = \dfrac{\sin^2 \beta}{\sin^2 \left(2 \alpha - \beta \right)}\left[ \dfrac{\sin^3 \alpha - \sin^3 \left(2 \alpha - \beta \right)}{\sin^3 \alpha - \sin^3 \beta} \right]  .
\end{equation}

Eqs.(\ref{anglestrap}) and (\ref{massratiotrap}) determine the isosceles trapezium central configuration. If we choose a mass ratio $m_2/m_1$, they are a system of two equations for the two unknown angles $\alpha, \beta$. As in the case of the kite (section \ref{sec:isosceles}), the direct problem (find the shape for given masses) is harder than the inverse problem (finding the masses for a given shape). However, the fact that there is only one mass ration here, rather than two as in the kite, makes the problem more tractable and indeed, for the isosceles trapezium there is a proof of existence and uniqueness for a given set of masses, by \citet{Xie}. Non-isosceles trapeziums also exist and have been investigated by, e.g., \citet{Santoprete a} and\citet{Corbera et. al.}, with \citet{Santoprete b} finally proving that there is at most one trapezoidal central configuration for any cyclic ordering of a given set of masses.

Here we only consider the simpler inverse problem. Eqs.(\ref{anglestrap}) and (\ref{massratiotrap}) imply that the value of, e.g., $\alpha$ completely determines the configuration, because once $\alpha$ has been chosen, Eq.(\ref{anglestrap}) determines the value of $\beta$. Given the two angles, Eq.(\ref{massratiotrap}) directly determines the masses needed to make the configuration central. 

Not every choice of $\alpha$ is allowed. Let us find the region of angles that $\alpha$ and $\beta$ can occupy. Because the ratio of masses is positive, we must have that 
\begin{equation*}
\dfrac{\sin^3 \alpha - \sin^3 \left(2 \alpha - \beta \right)}{\sin^3 \alpha - \sin^3 \beta} > 0
\end{equation*}
Since $\dfrac{\pi}{2} > \alpha > \beta$, the denominator is positive, and so must be the numerator, i.e., $\sin \alpha > \sin \left(2 \alpha - \beta \right)$. Because $\alpha < 2 \alpha - \beta$, the only solution is 
\begin{equation}
	2 \alpha - \beta > \pi - \alpha  \qquad \Longrightarrow  \qquad 3 \alpha - \pi > \beta >0
\end{equation}
In particular, it implies that $\alpha > \dfrac{\pi}{3}$.

In Eq.(\ref{anglestrap}), the left hand side is positive since it appears in the mass ratio. Thus, the right hand side must also be. We have two possible cases. Either the numerator and denominator are both negative, or they are both positive. Let us prove that the first case is impossible.

Assume that 
\begin{subequations}
	\begin{align*}
		\sin \left(2 \alpha - \beta \right) &< \sin \left(\alpha - \beta \right)  \\
		\sin \left(\alpha - \beta \right) &< \sin \beta
	\end{align*}
\end{subequations}
Since $\alpha < \dfrac{\pi}{2}$ and $2\alpha - \beta > \alpha - \beta$, the first equation can only hold if 
\begin{equation*}
	\pi > 2 \alpha - \beta > \pi - \left(\alpha - \beta \right) \qquad \Longrightarrow  \alpha > \dfrac{\pi + 2 \beta}{3}
\end{equation*}
Now since $\alpha < \dfrac{\pi}{2}$, this also puts a limit on $\beta$, such that $\beta < \dfrac{\pi}{4}$. This contradicts the second equation, however, according to which we must have that 
\begin{equation*}
 \pi - \left(\alpha - \beta \right) > \beta > \alpha - \beta
\end{equation*}
From here we have that $2\beta > \alpha$ but since $\alpha > \dfrac{\pi + 2 \beta}{3}$, this implies that $\beta > \dfrac{\pi}{4}$ which is a contradiction.

Hence, the numerator and denominator of the right hand side of Eq.(\ref{anglestrap}) must both be positive, i.e., 
\begin{subequations}
	\label{alphabeta2}
	\begin{align}
		\sin \left(2 \alpha - \beta \right) &> \sin \left(\alpha - \beta \right)  \label{alphabeta2:a}\\
		\sin \left(\alpha - \beta \right) &> \sin \beta   \label{alphabeta2:b}
	\end{align}
\end{subequations}
Eq.(\ref{alphabeta2:a}) implies that
\begin{equation*}
	\pi - \left(\alpha - \beta\right) > 2 \alpha - \beta > \alpha - \beta
\end{equation*}
The rightmost inequality is always fulfilled since $\alpha > 0$. Thus, the required condition is 
\begin{equation}
	\pi > 3 \alpha - 2 \beta
\end{equation}
Eq.(\ref{alphabeta2:b}) implies that
\begin{equation*}
	\pi - \beta > \alpha - \beta > \beta
\end{equation*}
or
\begin{equation}
	\alpha > 2 \beta
\end{equation}

\begin{figure}[h!]
	\includegraphics[width=0.7\linewidth]{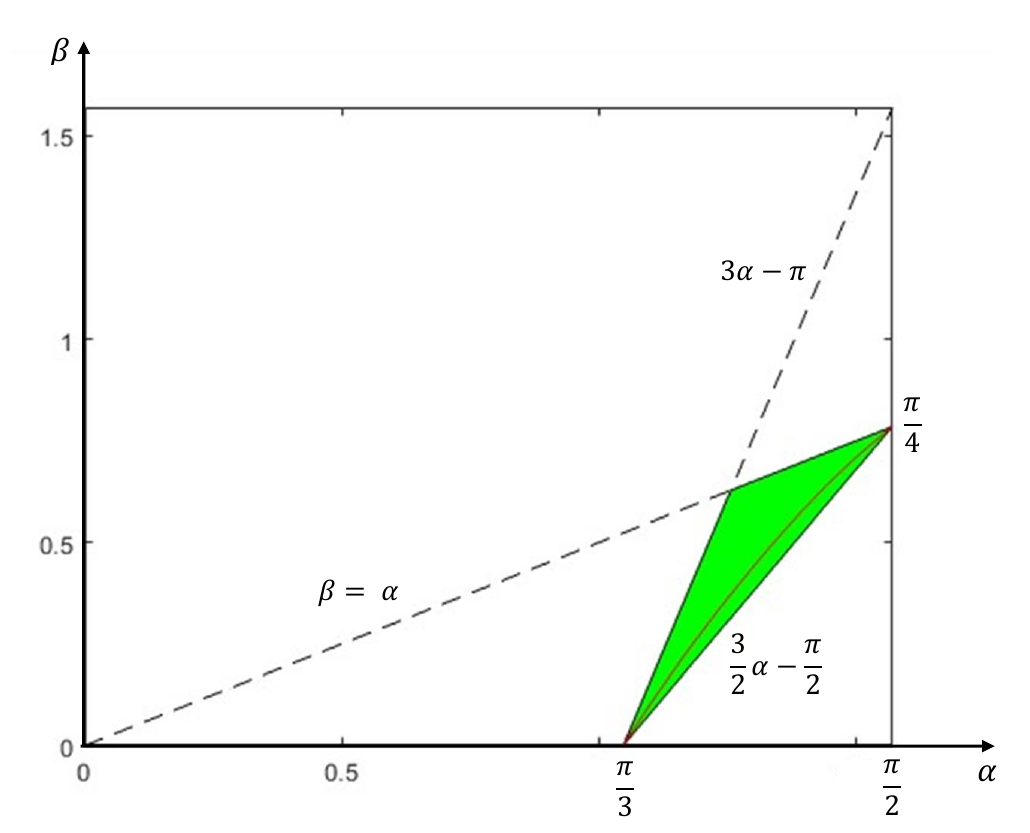}
	\caption{The possible angles for an isosceles trapezium central configuration. The allowed area is shaded in the graph. The solid line inside the shaded area is the function $\beta (\alpha)$ that gives the angle $\beta$ for every choice of base angle $\alpha$. There is a unique allowed isosceles trapezium for every base angle in the range $\left(\dfrac{\pi}{3} , \dfrac{\pi}{2}\right)$.}
	\label{Fig:trapmass}
\end{figure}

Combining all the conditions yields that 
\begin{subequations}
	\begin{align}
		\beta < \frac{\alpha}{2}   ,\\
		\beta < 3 \alpha -\pi     ,\\
		\beta > \dfrac{3 \alpha - \pi}{2}   .
	\end{align}
\end{subequations}
The various lines and their intersection are shown in Fig.\ref{Fig:trapmass}, which depicts the possible values of $\alpha$ and $\beta$ plotted against each other. The triangular shaded area is the region verifying all the limitations. 

However, Eq.(\ref{anglestrap}) determines a single value of $\beta$ for any base angle $\alpha$ (within the allowed range $\dfrac{\pi}{2} > \alpha > \dfrac{\pi}{3}$). Therefore there is a specific curve in the allowed region that describes the function $\beta \left( \alpha \right)$ that solves Eq.(\ref{anglestrap}). The numerical solution of that function is depicted in Fig.\ref{Fig:trapmass} as a solid curve within the shaded region. 

To summarize, once $\alpha$ is chosen, $\beta$ is determined from Eq.(\ref{anglestrap}), and the mass ratio $m_1/m_2$ is determined from Eq.(\ref{massratiotrap}). Alternatively, choosing a mass ratio produces a system of two equations, Eq.(\ref{anglestrap}) and Eq.(\ref{massratiotrap}), that needs to be solved in order to find the two angles $\alpha$ and $\beta$. These determine the isosceles trapezium's shape, up to scaling.

\section{Summary}

The present work continues the exploration of the advantages of the pair space representation in the N-body problem, by considering central configurations in the four body problem. Eqs.(\ref{4cc}) are the fundamental necessary and sufficient relations for a non-collinear configuration to be central (collinear configurations have been treated generally in a previous paper \citep{Drory b}). The form of the equations suggests classifying the configurations by the number of pair-distances that are equal.

\begin{enumerate}
\item If all $q_{ij}$'s are equal, the configuration is a tetrahedron. This is the only nonplanar four-body central configuration. 

\item If three masses form an equilateral triangle (three equal pair-distances), their masses must be equal. The fourth mass is arbitrary but must sit at the centre of the triangle.

\item If three masses form an isosceles triangle (two equal pair-distances), the configuration must be a kite. The masses at the end of the line perpendicular to the central axis of symmetry must be equal. The other two masses may be different. Their values are then mathematically related to the angles $\alpha, \beta$ of the kite. These relations limit the possible values of the angles to a definite domain in the $\alpha\beta$ plane. The shape of this domain differs according to the kite being convex or concave. The particular case $\alpha = \beta$ corresponds to a rhombus. This configuration must have two pairs of equal masses, located opposite one another. Up to scaling, the rhombus is determined by a single angle, which is the solution of an equation that depends on the ratio of the unequal masses.

\item If no triangle in the configuration has two equal sides, there may still be pairs of relative distances that are equal. If two such pairs exist, we have only one new solution, which is an isosceles trapezium (the rhombus shows up again as the only possible parallelogram solution, but it has been accounted for in the previous case already). The two masses at the end of the long base must be equal, and so are the two masses at the ends of the short base. Once again, the angles that define the configuration up to scaling are mathematically related to the ratio of the unequal masses.
\end{enumerate}

The pair-space approach describes central configurations in a purely algebraic fashion, as opposed to the standard analytic characterization as critical points of the potential under some constraint \citep{Moeckel}. This allows a fully vectorial treatment that is thus more general than the usual approaches, in which one would first choose a geometry and then fit to it the coordinates system. In particular, note that in the present treatment, no coordinate system is assumed at any stage. The various angles arise naturally from either scalar or cross products of vectors. Thus, there is no need to decompose the equations in a particular coordinate system. Consequently, the enumeration and analysis of the configurations proceeds in a more methodical, systematic and exhaustive fashion, compared to previous approaches. Here, every configuration with two or more pairs of equal relative distances has been counted and described. Thus, additional central configurations must fall in one of two classes:

\begin{enumerate}[label=(\alph*)]
\item All relative distances differ from one another.
\item If there is one pair of equal distances, it is the only one, and these distances cannot share a common vertex.
\end{enumerate}

The exploration of such additional configurations is not pursued here for lack of space and should be the subject of further research.

\section*{Declarations}

\textbf{Competing interests:} 

The author declares no competing interests.

\textbf{Data availability:} 

No datasets were generated or analysed during the current study.

\textbf{Research funding:}

This research did not receive any funding.

\textbf{Acknowledgments:} 

I am deeply grateful to an anonymous reviewer whose generous and very thorough remarks have immeasurably improved this manuscript.

\section{Appendix: Proof of existence of the body-centered equilateral triangle configuration}
\label{sec:proof}
\renewcommand{\theequation}{A.\arabic{equation}}
\setcounter{equation}{0}
\renewcommand{\thefigure}{A.\arabic{figure}}
\setcounter{figure}{0}

We want to show that the configuration described in section \ref{sec:equilateral}, i.e., the body-centered equilateral triangle, is in fact realizable; in other words, we want to prove that there exists a solution of the equations of motions that remains a body-centered equilateral triangle at all times, with only changes in scale or orientation allowed. The shape of the configuration is described in Fig.{\ref{Fig:4body_equilateral}.
	
Simple geometry shows that $q_{4 1} = q_{4 2} = q_{4 3} = \dfrac{1}{\sqrt{3}} q_{1 2}$. Furthermore, we have that
\begin{equation}
	\label{q4s}
	\bm{q}_{1 4} + \bm{q}_{2 4} + \bm{q}_{3 4} = 0   ,
\end{equation}
which derives from the fact that each of these vectors is the rotation by $\dfrac{2 \pi}{3}$ of its predecessor.

Let us now consider the equations of motions. These contain terms of the form $\bm{J}_{ij}$, expressed through functions $\bm{F}_{ijk}$ (see Eqs.(\ref{eqmotionq}), (\ref{jij}) and (\ref{eqmotionJ})). We shall now express these explicitly. Start with
\begin{equation}
	\bm{F}_{1 2 3} = GM\left[ \dfrac{\bm{q}_{1 2}}{q_{1 2}^3} + \dfrac{\bm{q}_{2 3}}{q_{2 3}^3} + \dfrac{\bm{q}_{3 1}}{q_{3 1}^3}\right] = \dfrac{G M}{q_{1 2}^3}\left[\bm{q}_{1 2} + \bm{q}_{2 3} +\bm{q}_{3 1}\right] = 0  .
\end{equation}
Next,
\begin{equation}
	\bm{F}_{1 2 4} = GM\left[ \dfrac{1}{q_{1 2}^3} \bm{q}_{1 2} + \dfrac{1}{q_{4 1}^3}\left(\bm{q}_{2 4} + \bm{q}_{4 1} \right)  \right] = \dfrac{G M}{q_{1 2}^3}\bm{q}_{1 2}\left[1 - \left( \sqrt{3}\right) ^3 \right]  ,
\end{equation}
where we have used the triangle condition $\bm{q}_{2 4} + \bm{q}_{4 1} = - \bm{q}_{1 2}$ and the ratio $q_{ 4 1} = \dfrac{1}{\sqrt{3}} q_{2 1}$.

By analogous calculations, we obtain:
\begin{subequations}
	\begin{align*}
		\bm{F}_{1 3 4} &= GM\left[ \dfrac{1}{q_{1 3}^3} \bm{q}_{1 3} + \dfrac{1}{q_{4 1}^3}\left(\bm{q}_{3 4} + \bm{q}_{4 1} \right)  \right] = \dfrac{G M}{q_{1 3}^3}\bm{q}_{1 3}\left[1 - \left( \sqrt{3}\right)^3 \right]  ,  \\
		\bm{F}_{2 3 4} &= GM\left[ \dfrac{1}{q_{2 3}^3} \bm{q}_{1 2} + \dfrac{1}{q_{4 2}^3}\left(\bm{q}_{3 4} + \bm{q}_{4 2} \right)  \right] = \dfrac{G M}{q_{2 3}^3}\bm{q}_{2 3}\left[1 - \left( \sqrt{3}\right)^3 \right]  ,
	\end{align*}
\end{subequations}

Substituting these expressions into Eq.(\ref{eqmotionJ}), we find
\begin{subequations}
	\begin{align}
		\dfrac{1}{\mu_{1 2}} \bm{J}_{1 2} &= m_3\bm{F}_{1 2 3} + m_4\bm{F}_{1 2 4}= \dfrac{G m_4}{q_{1 2}^3}\bm{q}_{1 2}\left[1 - \left( \sqrt{3}\right)^3 \right]  ,\\
		\dfrac{1}{\mu_{1 3}} \bm{J}_{1 3} &= m_2\bm{F}_{1 3 2} + m_4\bm{F}_{1 3 4}= \dfrac{G m_4}{q_{1 3}^3}\bm{q}_{1 3}\left[1 - \left( \sqrt{3}\right)^3 \right]  ,\\ 
		\dfrac{1}{\mu_{2 3}} \bm{J}_{2 3} &= m_1\bm{F}_{2 3 1} + m_4\bm{F}_{2 3 4}= \dfrac{G m_4}{q_{2 3}^3}\bm{q}_{2 3}\left[1 - \left( \sqrt{3}\right)^3 \right]  . 
	\end{align}
\end{subequations}

We shall now solve for the motion of the masses forming the equilateral triangle. This is determined by the pair-positions $\bm{q}_{12}, \bm{q}_{23}, \bm{q}_{13}$. With the condition $m_1 = m_2 = m_3$, these first three equations of motions are, from Eq.(\ref{eqmotionq}),
\begin{subequations}
	\label{4b123}
	\begin{align}
		\bm{\ddot{q}}_{12}  +  \frac{G \left( 3 m_1 + 3^{3/2} m_4 \right)}{q_{12}^3}\bm{q}_{12} &= 0   ,\label{4b123:a}   \\
		\bm{\ddot{q}}_{13}  +  \frac{G \left( 3 m_1 + 3^{3/2} m_4 \right)}{q_{13}^3}\bm{q}_{13} &= 0   ,\label{4b123:b}   \\
		\bm{\ddot{q}}_{23}  +  \frac{G \left( 3 m_1 + 3^{3/2} m_4 \right)}{q_{23}^3}\bm{q}_{23} &= 0   .\label{4b123:c}
	\end{align}
\end{subequations}

These are Keplerian equations of motions, and we can now write down their solution explicitly. Start by solving the equation of, e.g., $\bm{q}_{1 2}$ with appropriate initial conditions. Since at $t = 0$ we assume that the $(1,2,3)$ triangle is equilateral, we must have constant rotation matrices $\mathfrak{R}_{i j}$ defined for the indices $(i,j) \in \left( 1 , 2 , 3 \right)$:
\begin{equation}
	\bm{q}_{i j}(t = 0) = \mathfrak{R}_{i j} \bm{q}_{1 2} (t = 0)   .
\end{equation}  
The solutions of Eqs.(\ref{4b123}) are now
\begin{equation}
	\label{resqij}
	\bm{q}_{i j}(t) = \mathfrak{R}_{i j} \bm{q}_{1 2} (t)  ,
\end{equation}
as follows from the linearity of the equations, since Eq.(\ref{4b123:b}) is just Eq.(\ref{4b123:a}) multiplied by $\mathfrak{R}_{1 3}$ and Eq.(\ref{4b123:c}) is Eq.(\ref{4b123:a}) multiplied by $\mathfrak{R}_{2 3}$. This completes the solution of the first three pair-positions $\bm{q}_{12}, \bm{q}_{23}, \bm{q}_{13}$.

We now solve for the motion of the pair-positions describing the motions of the exterior three masses with respect to the central mass $m_4$. For these, the remaining three equations of motions, we need the expressions:
\begin{subequations}
	\label{j4i}
	\begin{align}
		\dfrac{1}{\mu_{1 4}} \bm{J}_{1 4} &= m_2\bm{F}_{1 4 2} + m_3\bm{F}_{1 4 3}= -\dfrac{G m_2}{q_{1 2}^3}\left[1 - \left( \sqrt{3}\right)^3 \right]\left(\bm{q}_{1 2} + \bm{q}_{1 3} \right)  ,\label{j4i:a}  \\
		\dfrac{1}{\mu_{2 4}} \bm{J}_{2 4} &= m_1\bm{F}_{2 4 1} + m_3\bm{F}_{2 4 3}= \dfrac{G m_1}{q_{1 2}^3}\left[1 - \left( \sqrt{3}\right)^3 \right]\left(\bm{q}_{1 2} - \bm{q}_{2 3} \right)    , \label{j4i:b}\\  
		\dfrac{1}{\mu_{3 4}} \bm{J}_{3 4} &= m_1\bm{F}_{3 4 1} + m_2\bm{F}_{3 4 2}= \dfrac{G m_1}{q_{1 2}^3}\left[1 - \left( \sqrt{3}\right)^3 \right]\left(\bm{q}_{1 3} + \bm{q}_{2 3} \right)    \label{j4i:c} ,
	\end{align}
\end{subequations}
where we have used the condition $m_1 = m_2 = m_3$ to relabel the masses. 

Using the triangle conditions and then Eq.(\ref{q4s}), we have that
\begin{subequations}
	\begin{align}
		\bm{q}_{12} + \bm{q}_{13} &= \left[ \bm{q}_{14} + \bm{q}_{42}\right]  + \left[ \bm{q}_{14} + \bm{q}_{43}\right]  = 3\bm{q}_{14} \\
		\bm{q}_{12} - \bm{q}_{23} &= \left[ \bm{q}_{14} + \bm{q}_{42}\right]  + \left[ \bm{q}_{42} + \bm{q}_{34}\right]  = - 3\bm{q}_{24}   \\
		\bm{q}_{13} + \bm{q}_{23} &= \left[ \bm{q}_{14} - \bm{q}_{34}\right]  + \left[ \bm{q}_{24} - \bm{q}_{34}\right]  = - 3\bm{q}_{34} .
	\end{align}
\end{subequations}
Hence, Eqs.(\ref{j4i}) become 
\begin{subequations}
	\label{j4ifinal}
	\begin{align}
		\dfrac{1}{\mu_{1 4}} \bm{J}_{1 4} &=-\dfrac{3 G m_1}{q_{1 2}^3}\left[1 - \left( \sqrt{3}\right)^3 \right]\bm{q}_{1 4}  ,\\
		\dfrac{1}{\mu_{2 4}} \bm{J}_{2 4} &=-\dfrac{3 G m_2}{q_{1 2}^3}\left[1 - \left( \sqrt{3}\right)^3 \right]\bm{q}_{2 4}  \\
		\dfrac{1}{\mu_{3 4}} \bm{J}_{3 4} &=-\dfrac{3 G m_3}{q_{1 2}^3}\left[1 - \left( \sqrt{3}\right)^3 \right]\bm{q}_{3 4}  ,
	\end{align}
\end{subequations}

Substitute this into the equations of motion of $\bm{q}_{i4}$ for any $i \neq 4$. From Eq.(\ref{eqmotionq}), we obtain
\begin{equation}
	\label{eqmoqi4}
	\bm{\ddot{q}}_{i4}  +  \frac{G M}{q_{i4}^3}\bm{q}_{i4} + \dfrac{3 G m_i}{q_{1 2}^3}\left[1 - \left( \sqrt{3}\right)^3 \right]\bm{q}_{i 4} = 0 ,
\end{equation}

At the moment $t = 0$, we can find matrices $\mathfrak{P}_{i}$ such that for $i \in \left(1 , 2, 3 \right)$ :
\begin{equation}
	\bm{q}_{i 4}(t = 0) =  \dfrac{1}{\sqrt{3}} \mathfrak{P}_{i} \bm{q}_{1 2} (t = 0)   .
\end{equation}
Then the explicit solutions of the equations are:
\begin{equation}
	\label{resqi4}
	\bm{q}_{i 4}(t) =  \dfrac{1}{\sqrt{3}} \mathfrak{P}_{i} \bm{q}_{1 2} (t)     .
\end{equation}

To see this, let us substitute these expressions into Eq.(\ref{eqmoqi4}). Because the equations are linear, the matrices 
$\mathfrak{P}_{i}$ are constant and $M - 3m_i = m_4$ for any $i \in (1 ,2 , 3)$, we find that the equations of motions become
\begin{equation}
	\dfrac{\mathfrak{P}_{i}}{\sqrt{3}} \left[ \bm{\ddot{q}}_{12}  +  \frac{G \left( 3 m_1 + 3^{3/2} m_4 \right)}{q_{12}^3}\bm{q}_{12}\right]  = 0
\end{equation}
The expression in square brackets is just the equation of motion of $\bm{q}_{12}$, i.e., Eq.(\ref{4b123:a}), and therefore the expressions in Eqs.(\ref{resqij}) and (\ref{resqi4}) are indeed a solution to the four-body problem that retains at all times the configuration of a body-centered equilateral triangle.

\end{document}